\documentclass[twocolumn]{aastex631}

\usepackage{amsmath}
\usepackage{subfigure}
\usepackage{tabularray}

\newcommand\revolver{\texttt{REVOLVER}}

\begin{document}

\title{Cosmic voids as a probe of the nature of dark matter: simulations and galaxy survey forecasts}

\author{Alexander Spencer London}
\affiliation{David A. Dunlap Department of Astronomy and Astrophysics, University of Toronto,\\
50 St George St, Toronto, ON, M5S 3H4, Canada}

\author{Keir K. Rogers}
\affiliation{Department of Physics, King's College London, Strand Building, Stand, London, WC2R 2LS, United Kingdom}
\affiliation{Department of Physics, Imperial College London, Blackett Laboratory, Prince Consort Road, London, SW7 2AZ, United Kingdom}
\affiliation{Dunlap Institute for Astronomy and Astrophysics, University of Toronto, 50 St George St, Toronto, ON, M5S 3H4, Canada}

\author[0000-0003-4642-6720]{Alex Laguë}
\affiliation{Department of Physics and Astronomy, University of
Pennsylvania,\\
209 South 33rd Street, Philadelphia, PA, 19104, United States of America}

\author[0000-0002-0965-7864]{Renée Hlo\v{z}ek}
\affiliation{Dunlap Institute for Astronomy and Astrophysics, University of Toronto, 50 St George St, Toronto, ON, M5S 3H4, Canada}
\affiliation{David A. Dunlap Department of Astronomy and Astrophysics, University of Toronto,\\
50 St George St, Toronto, ON, M5S 3H4, Canada}

\author{Zara Zaman}
\affiliation{Origins Institute, McMaster University, 1280 Main St W, Hamilton, ON, L8S 4M1, Canada}
\affiliation{David A. Dunlap Department of Astronomy and Astrophysics, University of Toronto,\\
50 St George St, Toronto, ON, M5S 3H4, Canada}
 
\begin{abstract}

Voids are parts of the cosmic web least affected by non-linearities and baryonic feedback. We thus calculate the sensitivity of voids to the nature of dark matter (DM), using ultra-light axions as a concrete model and the ongoing Dark Energy Spectroscopic Instrument (DESI) and \textit{Euclid} galaxy surveys as observational settings. We simulate axion effects on voids using mass-peak patch simulations and find that: (i) axions suppress the formation of lower-mass halos leading to the merging of smaller (radius \(< 25\,\mathrm{Mpc}/h\)) voids into fewer larger (radius \(> 25\,\mathrm{Mpc}/h\)) voids; and (ii) voids in the presence of axions are emptier of halos, thereby suppressing the void-halo correlation function. These effects strengthen as axion particle mass \(m_\mathrm{a}\) decreases. We forecast improvements in axion constraints from the void size function (VSF; the void number density as function of their radius). A \textit{Euclid}-like survey (effective volume of \(73\,\mathrm{Gpc}^3\) with a prior on the other \(\Lambda\)CDM cosmological parameters from the Simons Observatory cosmic microwave background experiment) can limit the axion energy density (for \(m_\mathrm{a} = 10^{-25}\,\mathrm{eV}\)) to \(< 4.6\%\) of the DM (at \(95\%\) credibility), about two times stronger than current limits. Conversely, we show that a Universe with a dark sector consisting of axions at the \(10\%\) level, as motivated by the string axiverse, can be recovered with \(\sim 2 \sigma\) preference. A DESI-like survey achieves comparable results. Axion and \(\Lambda\)CDM parameters have different degeneracies given VSF and galaxy power spectrum data, indicating future combined analyses will be most powerful in disentangling the DM nature. We anticipate our results will extend to other (e.g., warm or interacting) DM models.

\end{abstract}



\section{Introduction} \label{sec:intro}
The fundamental nature of dark matter is unknown. While direct detection of a dark matter particle has so far eluded us \citep[e.g.,][]{XENON:2023cxc,Butler:2023eah,LZ:2024zvo}, the large-scale structure of the Universe bears witness to its clustering properties \citep[e.g.,][]{Planck:2018vyg,DESI:2024mwx,AbdulKarim2025DESIDR2}. The clustering of overdensities is a well-established cosmological probe when testing the nature of dark matter \citep[e.g.,][]{Abazajian:2005xn,Lague:2021frh, Dentler:2021zij, Rogers:2023ezo}, but the properties of underdensities, i.e., voids \citep[e.g.,][]{2010MNRAS.403.1392L,Yang2015WarmthElevating,Nadathur:2015qua,Gallagher:2022zar}, have relatively recently become competitive probes.

Different dark matter models predict distinct clustering properties in both over- and underdensities. As observations of clustering statistics improve, so does our ability to discern between different particle models, if we build sufficiently accurate models of clustering for non-standard dark matter. The model on which we focus here is the axion (indeed, our results apply to any light scalar or pseudoscalar field with axions as a particular particle physics scenario). The axion is a dark matter particle candidate originally postulated to solve the strong CP problem in the Standard Model of particle physics \citep{PhysRevLett.38.1440,PhysRevLett.40.223,PhysRevLett.40.279} (for axions with a mass $m_\mathrm{a} \sim 10^{-5}\,\mathrm{eV}$). When axions are generated within high-energy theories, e.g., string theory \citep{WITTEN1984351,Svrcek:2006yi,PhysRevD.81.123530,Antypas:2022asj,Sheridan:2024vtt,Jain:2025vfh}, a population of ultra-light axion (ULA) particles (defined as \(m_\mathrm{a} < 10^{-17}\,\mathrm{eV}\)) is often produced, which has a distinct impact on cosmic clustering.

Besides providing a physical explanation for what dark matter could be, axions have also been proposed to solve some of the inconsistencies between simulations of a universe that include only cold, collisionless dark matter (CDM) and smaller-scale (sub-galactic) observations \citep[][]{Bullock:2017xww}. More complete observations of the Milky Way satellite galaxy population \citep{DES:2020fxi} and the effects of bursty supernovae feedback \citep{Pontzen:2014lma} have relieved much of these inconsistencies, although there may be more diversity in dwarf galaxy rotation curves than expected \citep{Oman:2015xda,2025arXiv251011800C}, which could be addressed by multi-field axion dark matter \citep{Luu:2024lfq}. ULAs are excellent dark matter candidates, being CDM-like on larger scales, while suppressing structure growth on scales below their astrophysically-sized \textit{de Broglie} wavelengths \citep[\(\sim\) kpc to tens of Mpc;][]{Hu:2000ke,Hlozek2015ASearch}. The Jeans suppression length \(\lambda_\mathrm{J} \propto m_\mathrm{a}^{-\frac{1}{2}}\), meaning that lighter axions affect larger scales.

ULAs with \(m_\mathrm{a} \sim 10^{-25}\,\mathrm{eV}\) improve consistency between cosmic microwave background (CMB) and galaxy surveys with regards to the \((1.5 - 3) \sigma\) discrepancy \citep[e.g.,][]{AbdulKarim2025DESIDR2,DES:2026fyc} in their inference of the ``clumpiness'' parameter \(S_8 \equiv \sigma_8 \left(\frac{\Omega_\mathrm{m}}{0.3}\right)^\frac{1}2{}\) \citep[][]{Rogers:2023ezo}, where $\sigma_8$ is the root-mean-squared linear density fluctuation on scales of $8\;\mathrm{Mpc}/h$ and $\Omega_\mathrm{m}$ is the total matter energy density. Incorrect modeling of baryonic feedback may also explain this discrepancy \citep{Amon:2022azi}. A tension has emerged when comparing CMB, baryon acoustic oscillations and supernovae observations with the Lyman-$\alpha$ forest as measured by the extended Baryon Oscillation Spectroscopic Survey \citep[eBOSS;][]{2019JCAP...07..017C,Fernandez:2023grg,Walther:2024tcj}, which can also be addressed by a population of ULAs with \(m_\mathrm{a} \sim 10^{-25}\,\mathrm{eV}\) \citep[][]{Rogers:2023upm}. Dark Energy Spectroscopic Instrument Lyman-$\alpha$ forest results \citep{Chaves-Montero:2026hqd} are consistent with CMB data thanks to improved modeling of metals \citep{Karacayli:2023oqv} and high-density absorbers \citep{Rogers:2017bmq,Rogers:2017eji}.

Hints of parity-violating cosmic birefringence can be sourced by ULAs with \(m_\mathrm{a} \sim 10^{-28}\,\mathrm{eV}\) \citep[e.g.,][notwithstanding instrumental and astrophysical systematics]{Minami:2020odp,PhysRevD.106.063503,Diego-Palazuelos:2022cnh,Gasparotto:2023psh,Luo:2023cxo,Yin:2024fez,Zhang:2024dmi}. Considerations both from fundamental theory and the cosmological model therefore motivate a search for ULAs across a range of particle masses. Since a ULA of a certain mass may not constitute all of the dark matter, this motivates a search across a range of cosmic energy densities (or fractions of the total dark matter).

The leading order effect of a ULA component is a partial scale-dependent suppression of the linear matter power spectrum relative to the pure CDM limit \citep{Hu:2000ke,Amendola:2005ad,Marsh:2010wq,Hlozek2015ASearch,Hui:2016ltb}. The wavenumber of the suppression increases with \(m_\mathrm{a}\) and the strength of the suppression increases with the axion fraction \(\Omega_\mathrm{a}/\Omega_\mathrm{d}\), where \(\Omega_\mathrm{d}\) is the total dark matter energy density comprised of both axions and a cold component and \(\Omega_\mathrm{a}\) is only the axion energy density. The CMB and the mildly non-linear clustering of galaxies bound ULAs for \(m_\mathrm{a} \leq 10^{-26}\,\mathrm{eV}\) to be less than \((1 - 10)\) \% of the dark matter \citep{Lague:2021frh,Rogers:2023ezo,AtacamaCosmologyTelescope:2025nti,Verdiani:2025jcf}. Constraints on the axion density at larger masses (\(m_\mathrm{a} \geq 10^{-25}\,\mathrm{eV}\)) are more non-trivial since the equivalent suppression scale enters the fully non-linear regime meaning that axion non-linearities must be calculated \citep{Dentler:2021zij,Vogt:2022bwy,Winch:2024mrt,Preston:2025tyl}.

The effects of axions in non-linear structure formation is computed through computationally-expensive cosmological simulations \citep{Schive:2014dra,PhysRevLett.123.141301,May:2021wwp,Lague:2023wes} and semi-analytical approaches like the halo model \citep{Marsh:2016vgj,Dentler:2021zij,Vogt:2022bwy,Dome:2024hzq,Winch:2024mrt}. However, it is powerful to consider, in addition, probes of the large-scale structure that remain quasi-linear to smaller scales, leaving a clearer signature of the linear suppression caused by ULAs. The Lyman-$\alpha$ forest is one such probe since it is sourced by intergalactic gas at about mean cosmic density and so is much less affected by non-linearities than galaxy probes at equivalent scales which trace the most dense regions of the dark matter \citep{Croft:1997jf,Lukic:2014gqa,Rogers:2020ltq,Rogers:2020cup,Rogers:2021byl,Irsic:2023equ}. However, the Lyman-$\alpha$ forest can only be observed from the ground for redshifts \(z > 2\) (and the signal gets weaker for \(z < 2\) in any case). The processes of hydrogen and helium reionization must also be carefully modeled in order to set robust cosmological constraints \citep{Molaro:2021tdz,Bird:2023evb} from this probe. In order to trace linear perturbations to low redshift (\(z < 2\)), the most underdense regions of the dark matter (voids) are ideal.

Voids are already identified as a powerful probe of, e.g., modified theories of gravity \citep[e.g.,][]{Clampitt:2012ub,Cai:2014fma,PhysRevD.95.024018,Perico:2019obq,Takadera:2025ehm} and the cosmic neutrino density \citep[e.g.,][]{2015JCAP...11..018M}, as well as a practical way to extract more information from galaxy surveys beyond the two-point correlation function of density peaks \citep[halos; e.g.,][]{Hamaus:2020cbu}. In this work, we explore the potential of voids as a probe of the nature of dark matter through deviations from the standard CDM model. We hypothesize that voids will add information since they will be sensitive to any dark matter effect on the matter power spectrum, while being a cleaner probe of linear theory. Here, in our first study, we restrict ourselves to an analysis of dark matter-only simulations, looking at the mass and size distributions of halos and voids and the correlation functions between them. We discuss the prospects of observing the effects of axions and other dark matter models through two main probes: the void-halo cross-correlation and the distribution of void sizes as inferred from large-volume galaxy surveys.

In \S~\ref{sec:simulations}, we describe the simulations and, in \S~\ref{sec:halo-corr}, we present measurements of the halo mass function (HMF) and halo-halo correlation function from the simulations. In \S~\ref{sec:voidfinder}, we explain how we define and find voids. We present results of the void analysis in \S~\ref{sec:voidanalysis} and then carry out a forecast using the void size function in galaxy surveys in \S~\ref{sec:voidforecast}. We then discuss results in \S~\ref{sec:discussion} and conclude in \S~\ref{sec:conclusion}.

\section{Mass-peak patch simulations}\label{sec:simulations}

We run a set of dark matter-only large-scale structure simulations to create halo catalogs. We use the mass-peak patch simulation code~\citep{Stein2019TheMassPeak} to create a suite of cubic boxes with a baseline volume of $(2048\;\mathrm{Mpc})^3$ and a baseline cell resolution of $(0.5\,\mathrm{Mpc})^3$ (i.e., $4096^3$ grid cells in total). The mass-peak patch algorithm allows us to generate large volumes at low computational costs compared to full \(N\)-body approaches. There are four main steps: (1) we generate a density field from a linear matter power spectrum using the adapted Boltzmann code \texttt{AxionCAMB} \citep{Hlozek2015ASearch}; (2) we find density peaks (defined as regions with an overdensity above a critical value $\delta_\mathrm{c} = 1.5$) and their Lagrangian radius; (3) we merge overlapping peaks (halos); and (4) we move the peak locations using second-order Lagrangian perturbation theory to their final positions and define halos by these peaks. The final output is a halo catalog at \(z = 0\) with positions, masses and radii. We define voids by locating local number underdensities of halos (see \S~\ref{sec:voidfinder} for a full description of the void finding procedure).

The simulations have as input fiducial \(\Lambda\)CDM parameters $[\sigma_8=0.8,\,\Omega_\mathrm{m}=0.31,\,n_\mathrm{s}=0.9655,\,\Omega_\mathrm{b}h^2=0.02222,\,H_0=67.37\,\mathrm{km/s/Mpc}]$, where $n_\mathrm{s}$ is the spectral index of the primordial power spectrum, $\Omega_\mathrm{b}h^2$ is the physical baryon energy density, $H_0$ is the Hubble constant and $h=H_0/(100\,\mathrm{km/s/Mpc})$.

\begin{table}
\centering
\begin{tabular}{|c|c|c|c|c|}
\hline
Sim. \# & Volume \(\left(\mathrm{Mpc}^3\right)\) & \# cells & \(\Omega_\mathrm{a} / \Omega_\mathrm{d}\) & \(m_\mathrm{a}\,(\mathrm{eV})\)\\ \hline
1              & \(512^3\)   & \(1024^3\)       & 0            & \(\Lambda\)CDM                        \\ 
2                & \(1024^3\)  & \(2048^3\)       & 0            & \(\Lambda\)CDM                        \\ 
3                & \(2048^3\)  & \(4096^3\)       & 0            & \(\Lambda\)CDM                        \\ \hline
4                & \(512^3\)   & \(1024^3\)       & 0.1           & $10^{-25}$  \\ 
5                & \(1024^3\)  & \(2048^3\)       & 0.1           & $10^{-25}$  \\ 
\textbf{6}                & \(\mathbf{2048^3}\)  & \(\mathbf{4096^3}\)       & \textbf{0.1}           & $\mathbf{10^{-25}}$  \\ \hline
7                & \(2048^3\)  & \(4096^3\)       & 0.1           & $10^{-23}$ \\ 
8                & \(2048^3\) & \(4096^3\)       & 0.1           & $10^{-24}$  \\ 
\textbf{6}                & \(\mathbf{2048^3}\)  & \(\mathbf{4096^3}\)       & \textbf{0.1}           & $\mathbf{10^{-25}}$  \\ 
9                & \(2048^3\)  & \(4096^3\)       & 0.1           & $10^{-26}$  \\ \hline
10                & \(2048^3\)  & \(512^3\)       & 0.1           & $10^{-25}$  \\ 
11                & \(2048^3\)  & \(1024^3\)       & 0.1           & $10^{-25}$  \\ 
12                & \(2048^3\)  & \(2048^3\)       & 0.1           & $10^{-25}$  \\ 
\textbf{6}                & \(\mathbf{2048^3}\)  & \(\mathbf{4096^3}\)       & \textbf{0.1}           & $\mathbf{10^{-25}}$  \\ 
\hline
\end{tabular}
\caption{Numerical and ULA parameters of the mass-peak patch simulations that we run. The volume (simulation \#1-3 for \(\Lambda\)CDM and \#4-6 for axion mass \(m_\mathrm{a}=10^{-25}\,\mathrm{eV}\)) and cell resolution (simulation \#10-12 and \#6) convergence tests are described in Appendix~\ref{app:simulation}. Simulation \#6 (\(m_a=10^{-25}\,\mathrm{eV}\)) with the largest volume and resolution defines our baseline study.}
\label{parameters-table}
\end{table}

\begin{figure*}
    \centering
    \includegraphics[width=0.9\linewidth]{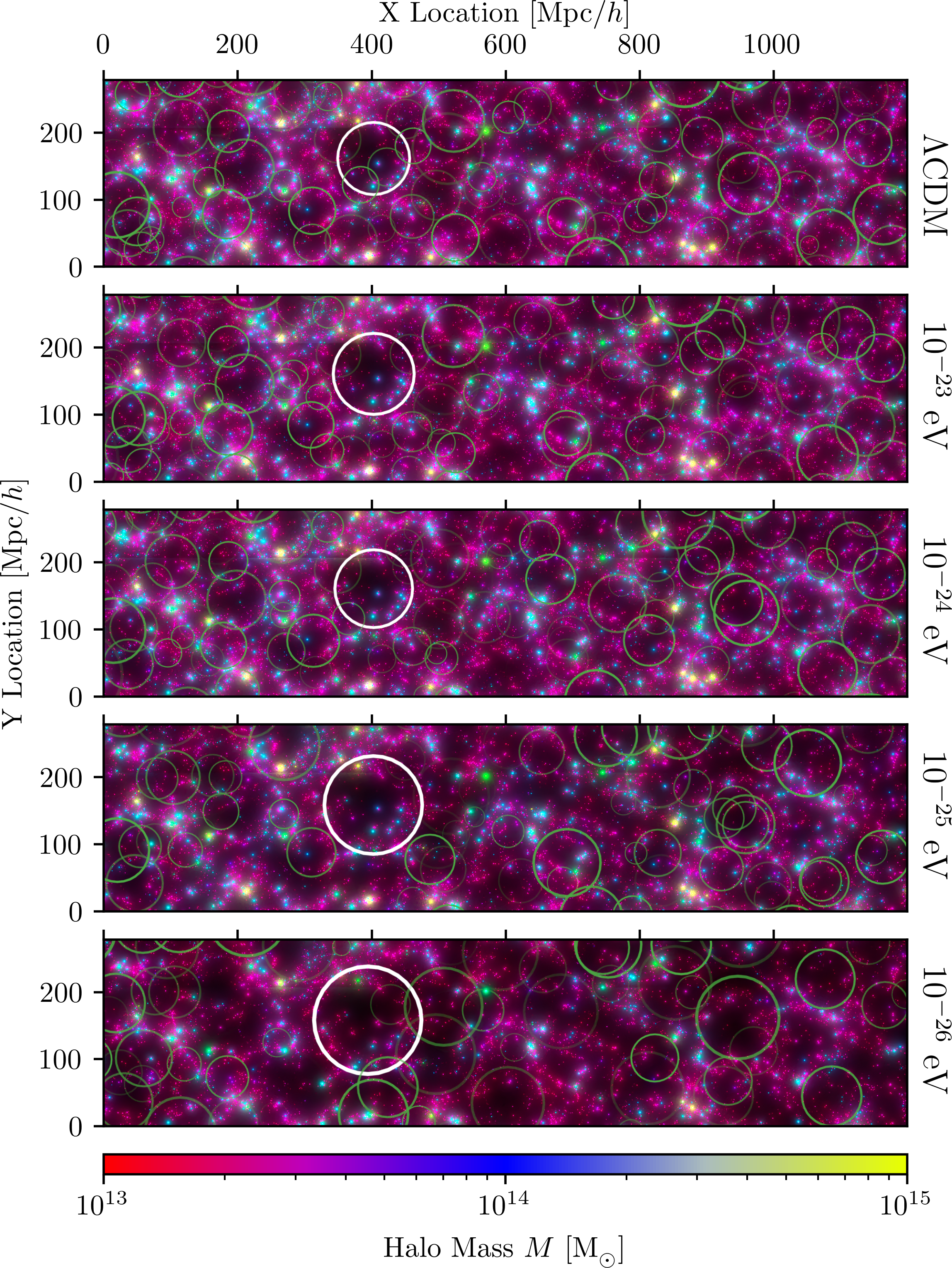}
    \caption{Voids (their positions and radii indicated by open green circles) identified (see \S~\ref{sec:voidfinder}) in slices of our halo simulations (\#3 and \#6-9, see Table \ref{parameters-table}) with the same random seed for the $\Lambda$CDM setting (\textit{top}) and axion masses decreasing from  $10^{-23}\,\mathrm{eV}$ to $10^{-26}\,\mathrm{eV}$ (\textit{top to bottom}). Halos and their masses are indicated by the colored points. The slices are \(50\,\mathrm{Mpc}/h\) thick in the Z direction (into the page). The brightness of each circle increases as the center of the void is closer to the center of the simulation slice in the Z direction. As axion mass decreases, there are on average fewer low-mass halos and voids increase in size. We highlight one such void in white.}
    \label{fig:Visualization}
\end{figure*}

We vary the ULA mass and dark matter fraction, as well as numerical properties of the simulations, as described in Table~\ref{parameters-table}. For simulation \#3, we fix the axion fraction to zero and this is the reference $\Lambda$CDM setting. We run the same \(\Lambda\)CDM model with different box volumes at fixed cell resolution in simulation \#1-3 and do the same volume convergence test for $m_\mathrm{a} = 10^{-25}\,\mathrm{eV}$ in simulation \#4-6. We also test convergence as we change the cell resolution at fixed box volume in simulation \#10-12 and \#6. We describe the numerical convergence tests in more detail in Appendix~\ref{app:simulation}. To test the effect of axion mass on void properties, in simulation \#6-9, we vary $10^{-26}\,\mathrm{eV} \leq m_\mathrm{a} \leq 10^{-23}\,\mathrm{eV}$, fixing the axion fraction to 10\% of the total dark matter density. In Fig.~\ref{fig:Visualization}, we show slices through the simulations as we vary axion mass. As the axion mass decreases, the Jeans suppression scale increases and so there are on average fewer low-mass halos, leaving larger voids. We discuss the effects on voids in more detail in \S~\ref{sec:voidfinder} and \ref{sec:voidanalysis}.

\begin{figure}
    \centering
\includegraphics[width=1\linewidth]{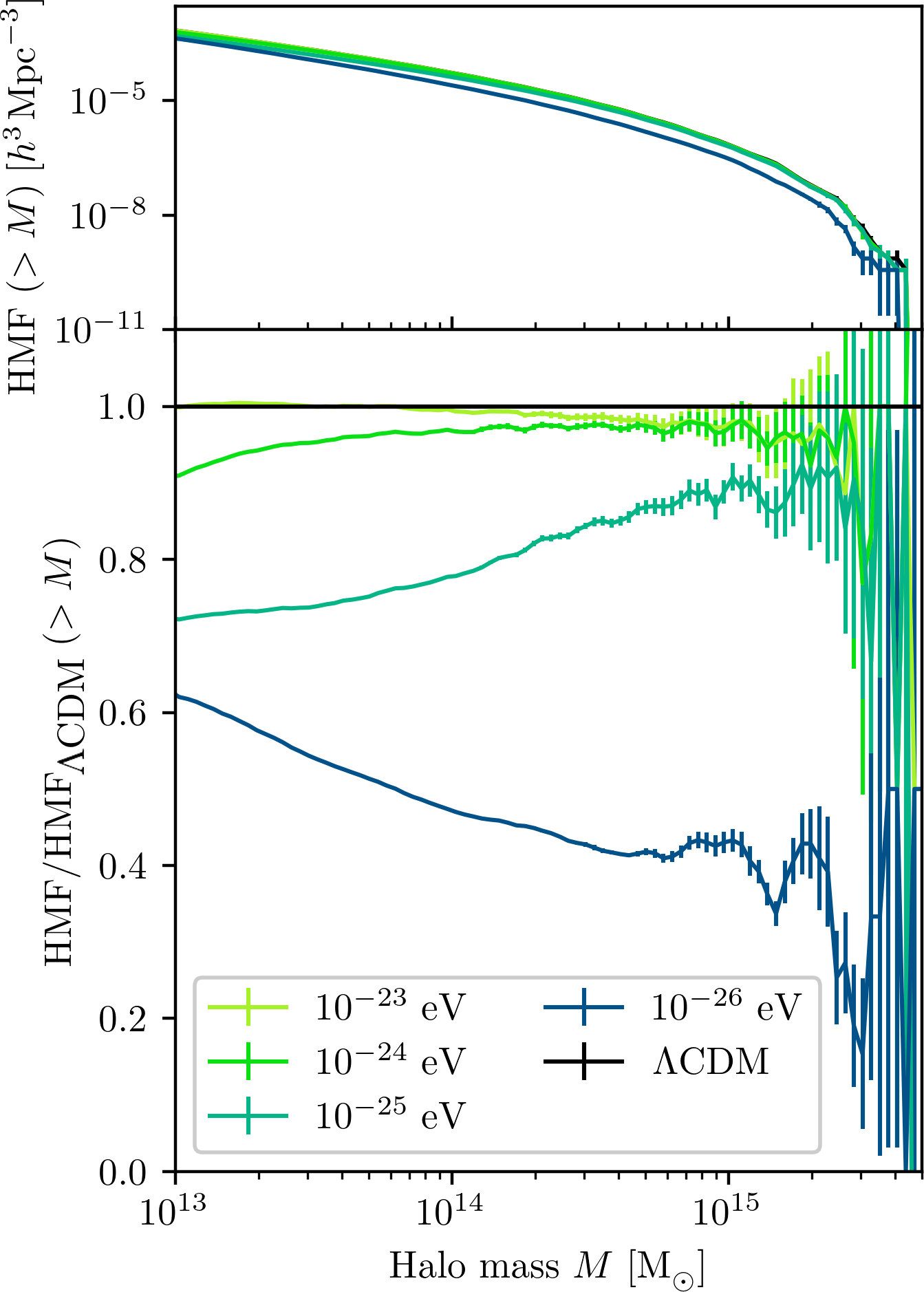}
    \caption{The halo mass function (HMF) of simulation \#3 and \#6-9 (see Table \ref{parameters-table}), showing the effect of decreasing axion mass. The \textit{upper} panel shows the cumulative HMF (number density of halos $>$ halo mass \(M\)). The \textit{bottom} panel shows the ratio of the HMF in the presence of axions to the $\Lambda$CDM HMF. We indicate the sample variance by the 68\% confidence error bars. As axion mass decreases, there are fewer low-mass halos.\label{fig:HMF1}}
\end{figure}

\section{Halo mass function and halo-halo correlation\label{sec:halo-corr}}
\subsection{Halo mass function} \label{sec:HMF}

The cumulative number density of halos above a certain mass, otherwise known as the halo mass function (HMF), measures how the hierarchical formation of halos depends on cosmological parameters. In Fig.~\ref{fig:HMF1}, we compare the HMF with varying axion mass to the \(\Lambda\)CDM case. As the axion mass decreases, its \textit{de Broglie} wavelength increases. Since axions do not cluster below this scale, the amplitude of initial density fluctuations is suppressed below this scale. This suppression has two main effects. First, the \textit{abundance} of low-mass halos is suppressed as these would form from smaller-scale perturbations that have themselves been suppressed. Second, the \textit{onset of structure formation} is delayed: more time is needed for density fluctuations to grow and reach the critical threshold in order to form a halo. When the axion mass reaches \(m_\mathrm{a} \sim 10^{-26}\,\mathrm{eV}\), this delay means that there is, in fact, a relative boost in the number density of low-mass halos compared to high-mass halos, as low-mass halos have not had time to merge into heavier objects \citep{Winch:2024mrt}. While these two effects are competing, nonetheless, the total number of halos is still always suppressed relative to the \(\Lambda\)CDM limit.

In order to make fair comparisons between models, we need to remove potential spurious halos (halos not well resolved by the limited spatial resolution of each simulation). We therefore remove from each simulation all halos with a mass below $10^{13}\; M_\odot$ (see Appendix \ref{app:simulation} for further details). This threshold is consistent with keeping only halos that contain at least a few hundred Lagrangian particles within their volume \citep{Stein2019TheMassPeak}. We now define the statistics used to analyse these simulations.

\subsection{Two-point correlation function}
\label{sec:2pcf}

The two-point correlation function of density contrasts $\delta_i(\mathbf{x}) = \rho_i(\mathbf{x})/\bar{\rho_i}-1$, where \(\rho_i(\mathbf{x})\) is the density of tracer \(i\) at position \(\mathbf{x}\) (and the bar over the quantity denotes an average over volume) is:
\begin{align}
    \xi_{ij}(\mathbf{x}-\mathbf{x}^\prime) = \left\langle\delta_i(\mathbf{x})\delta_j(\mathbf{x}^\prime) \right\rangle.
\end{align}
In this work, the tracers we consider are halos and voids. In the case of discrete tracers like these, the contrast is given in terms of number densities $n_i$ such that $\delta_i(\mathbf{x}) = n_i(\mathbf{x})/\bar{n_i} - 1$. Our simulations are isotropic random fields without redshift-space distortions and so the correlation function depends only on the magnitude of the separation \(r = |\mathbf{x} - \mathbf{x^\prime}|\). The correlation function of two fields is the Fourier transform of the cross power spectrum $P_{ij}(k)$:
\begin{align}
    \xi_{ij}(r) = \int \frac{\mathrm{d}k\,k^2}{2\pi^2 }  \frac{\sin kr}{kr} P_{ij}(k),
\end{align}
where $k$ is wavenumber. Thus, \(\xi_{ij} (r)\) is sensitive to axions through the Jeans suppression in \(P_{ij}(k)\). By measuring the correlation function in our simulations, we are thus probing the effects of axions on the \(z = 0\) matter power spectrum.

In practice, we compute the correlation function of the tracers (halos and voids) in the simulations using the Landy-Szalay estimator \citep{landyszalay}. This estimator of \(\xi_{ij}(r)\) counts how many pairs of tracers are separated by a distance $r$:
\begin{equation}
\hat{\xi}_{ij}(r) = \frac{(D_iD_j-D_iR_j-D_jR_i+R_iR_j)|_r}{R_iR_j|_r},\label{eq:landy}
\end{equation}
where $(D_iD_j)|_r$ is the number of pairs separated by a distance \(r\) between the $i^\mathrm{th}$ and $j^\mathrm{th}$ data catalogs \(D\) (i.e., the halo or void catalog from the mass-peak patch simulations), $(D_{i}R_{j})|_r$ is the pair count between the $i^\mathrm{th}$ data catalog and the $j^\mathrm{th}$ random catalog \(R\) and $(R_{i}R_{j})|_r$ is the pair count between the $i^\mathrm{th}$ and $j^\mathrm{th}$ random catalogs. The random catalogs are drawn from a spatially-uniform distribution, i.e., without the signal in the simulations. In other words, the Landy-Szalay estimator measures the excess in pairs of tracers separated by a distance \(r\) relative to a random uniform distribution as a measure of the ``clumpiness" of the tracer distributions.

The number of pairs between two objects in a set of $N$ objects is \(N(N-1)/2 \approx N^{2}/2\) for \(N \gg 1\). Thus, the computational cost of using the Landy-Szalay estimator is prohibitive for the large catalogs that we have on the order of $10^6$ halos. We instead employ the version of the estimator provided within the \texttt{corrfunc} code suite\footnote{\url{https://corrfunc.readthedocs.io}.} \citep{corrfunc}, which uses a tree method for quick pair counting, scaling as \(N \log N\).

\subsection{Jackknife errors}
\label{sec:errors}

In order to compute errors on the correlation, mass and size functions that we measure in this work (see \S~\ref{sec:halo-corr} and \ref{sec:voidanalysis}), we use jackknife resampling. Here, we divide each catalog into \(n_\mathrm{SS}\) independent sub-samples.\footnote{We sub-sample randomly from each halo and void catalog. We also tested sub-sampling by removing contiguous sub-volumes. We optimized \(n_\mathrm{SS} \in [4, 64]\), finding that for our catalogs \(n_\mathrm{SS} = 16\) achieved a balance between having many sub-samples for good covariance estimation (\(n_\mathrm{SS}\) not being too small) and varying the catalog sufficiently (\(n_\mathrm{SS}\) not being too large).} We then in turn remove each of the \(n_\mathrm{SS}\) sub-samples and recalculate the halo/void statistic. The covariance of the statistic between data bins \(a\) and \(b\) is then \citep[here shown specifically for the correlation function \(\xi\),][]{faizan/percival:2022}
\begin{equation}
    \mathcal{C}_{ab} = \frac{n_\mathrm{SS}-1}{n_\mathrm{SS}}\sum_{k=1}^{n_\mathrm{SS}}\left[\xi_{a}^{[k]} -\bar{\xi}_{a} \right]\left[\xi_{b}^{[k]} -\bar{\xi}_{b} \right],
\end{equation}
where the average over \(n_\mathrm{SS}\) jackknife realizations \(\xi^{[k]}\)
\begin{equation}
 \bar{\xi}_{a} = \frac{1}{n_\mathrm{SS}}\sum_{k=1}^{n_\mathrm{SS}}  \xi_{a}^{[k]}.
\end{equation}
Jackknife resampling accounts for sample variance but does not account for systematic uncertainties arising from the limits of volume and resolution in the numerical simulations. These uncertainties are determined in Appendix \ref{app:simulation}.

\begin{figure}
    \centering
    \includegraphics[width=1\linewidth]{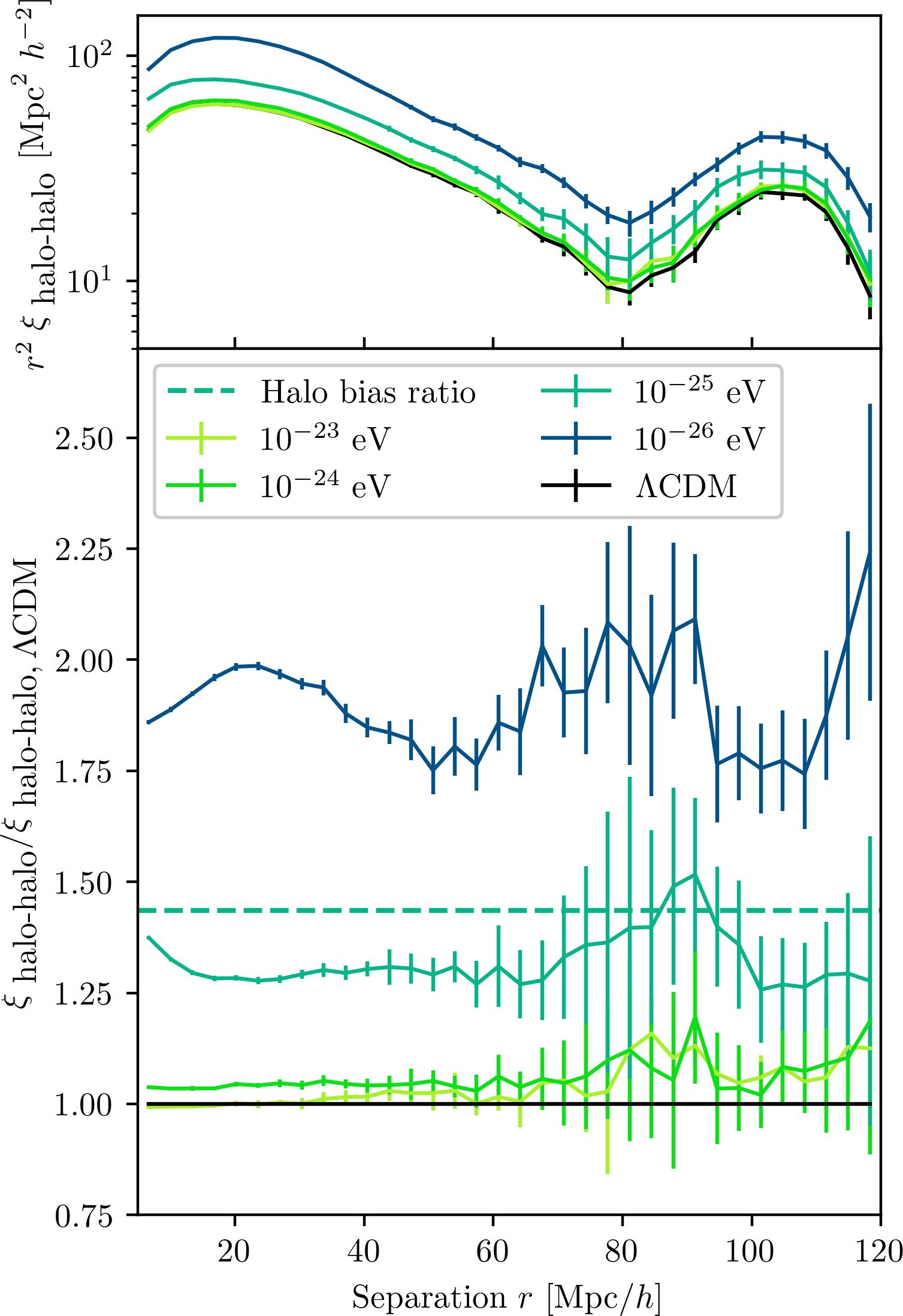}
    \caption{\textit{In the top panel}, the halo-halo correlation function \(\xi (r)\) as a function of separation \(r\) for simulation \# 3 and 6-9 (see Table \ref{parameters-table}), showing the effect of decreasing axion mass. \textit{In the bottom panel}, we show the ratio to the \(\Lambda\)CDM case. We indicate the sample variance by the 68\% confidence limit error bars. As axion mass decreases, the halo bias increases and so the correlation function increases. The dashed line indicates the halo bias ratio \(\left(\frac{b_{\mathrm{h},10^{-25}\,\mathrm{eV}}}{b_{\mathrm{h},\Lambda\mathrm{CDM}}}\right)^2\), which we will use in \S~\ref{sec:voidforecast}.\label{fig:halohaloCF}}
\end{figure}

\subsection{Axions lead to higher halo bias and boosted correlation function}
\label{sec:axion_halo_bias}

We show the correlation function between halos and its dependence on axion mass in Fig.~\ref{fig:halohaloCF}. The feature at \(r \sim 100\,\mathrm{Mpc}/h\) is the expected baryon acoustic oscillation feature, the scale of which is unaffected by these axion masses (it occurs at the same separation). However, we observe an increase in the amplitude of the halo-halo correlation function at all separations when decreasing the axion mass. As the axion mass decreases (at fixed axion fraction), the total number of halos is suppressed, but not uniformly (which is shown in Fig.~\ref{fig:HMF1}). Halos affected by axions are typically lower-mass halos which are more likely to form in lower-density regions, which are by definition the less-clustered regions. As a result, a lower axion mass results in a \textit{higher halo bias}, which is defined as the ratio of the halo overdensity to the matter overdensity on large, linear scales. In other words, the halo bias measures how much \textit{more} clustered halos are compared to the underlying linear matter field. While linear perturbations are suppressed, the higher-mass halos in more-clustered, higher-density regions are relatively less affected compared to the lower-mass halos in less-clustered regions, which leads to an increase in the halo bias. Since the halo correlation function is proportional to the square of the halo bias, this quantity in turn also increases. Therefore, even if the total matter power spectrum (or two-point correlation function) is dampened by ULAs, \textit{the correlation function of biased tracers (such as halos or galaxies) is boosted by this increase in bias} \citep[for the effect of ULAs on tracer bias, see also][]{Lague:2021frh}. The impact of axions on the \textit{void} correlation function is different (see \S~\ref{sec:voidanalysis}), leading to unique signatures when combining clustering of halos and voids.

In Fig.~\ref{fig:halohaloCF}, we show the halo bias ratio \(\left(\frac{b_{\mathrm{h},10^{-25}\,\mathrm{eV}}}{b_{\mathrm{h},\Lambda\mathrm{CDM}}}\right)^2 = \left(\frac{1.51}{1.26}\right)^2\), where \(b_{\mathrm{h, X}}\) is the halo bias in simulation X. We calculate the halo bias in each simulation by fitting a constant \(b_\mathrm{h}^2\) to the ratio of the halo-halo power spectrum to the linear matter power spectrum on large scales \citep{1984ApJ...284L...9K}. We will use these values in the void size function forecast in \S~\ref{sec:voidforecast}.

\section{Void finder} \label{sec:voidfinder}
For each halo catalog, we generate an accompanying catalog of voids. We use the REal-space VOid Locations from surVEy Reconstruction \citep[\revolver;][]{revolver,Nadathur2019BeyondBAO} code to find voids in each simulation. We use the ZOnes Bordering On Voidness \citep[\texttt{ZOBOV};][]{neyrinck} method. A void is a region of matter underdensity. Since our simulation forms a halo catalog (and real observations give the positions of biased tracers like galaxies), we define voids as an underdensity of the number of halos. \texttt{ZOBOV} identifies voids as a local minimum in the halo number density with a depression around it. There are three main steps in the algorithm: (1) we define cells around each halo by Voronoi tessellation to calculate the local halo number density; (2) we combine cells into zones using a watershed algorithm; and (3) we then define each unique zone as a void. These voids are not necessarily spherical, though we sometimes approximate them as such for some of our analysis. Following \citet{Nadathur2019BeyondBAO}, the position of a void is taken as the center of the largest sphere devoid of halos that fits within the void. We compute the mass of a void by summing individual halo masses contained within it. We compute the effective radius of the void as the radius of a sphere that has the same volume as the void. The \texttt{ZOBOV} hyper-parameter \texttt{guard\textunderscore nums = 30} controls the number of additional ``guard'' particles used to stabilize the tessellation near the edge of the simulation box. The parameter \texttt{zobov\textunderscore box\textunderscore div = 2} divides the Voronoi tessellation calculation into eight sub-volumes, each with about half the box length of the full simulation. The parameter \texttt{zobov\textunderscore buffer} is the fractional overlap between adjacent sub-volumes, tuned to 0.06 for a $(2048\,\mathrm{Mpc})^3$ box, 0.08 for a $(1024\,\mathrm{Mpc})^3$ box and 0.18 for a $(512\,\mathrm{Mpc})^3$ box.

\begin{figure}
    \centering
    \includegraphics[width=1\linewidth]{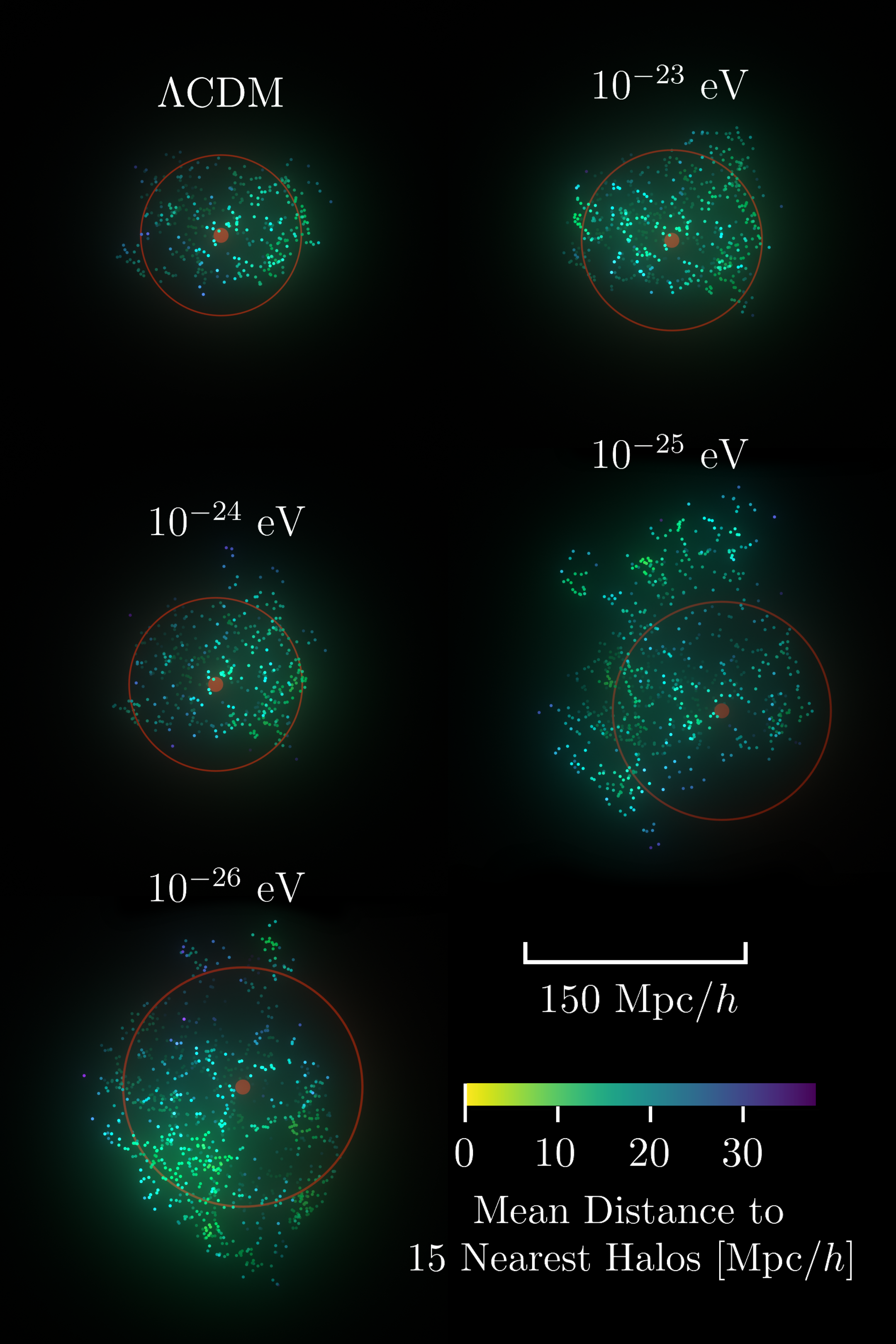}
    \caption{A single void (the same as highlighted in white in Fig.~\ref{fig:Visualization}) identified from our halo simulations (\#3 and \#6-9, see Table \ref{parameters-table} for details) with the same random seed for the \(\Lambda\)CDM setting (\textit{top left}) and axion masses decreasing from \(10^{-23}\,\mathrm{eV}\) to \(10^{-26}\,\mathrm{eV}\) (\textit{top to bottom, left to right}). Halos within the void and their local number density (specifically, the mean distance to the fifteen nearest halos) are indicated by the colored points. The center and radius of the void (defined in the main text) are respectively indicated by the red points and circles. As axion mass decreases, voids increase in size and get emptier of halos. \label{fig:single_void}}
\end{figure}

We illustrate the positions and sizes (by spherical approximation) of voids in Fig.~\ref{fig:Visualization}. As the axion mass decreases and there are on average fewer low-mass halos, this increases on average the size of depressions (or ``zones'') around local halo number density minima. In other words, voids on average increase in size; one such void is highlighted in white in Fig.~\ref{fig:Visualization}. We note that since the Voronoi tesselation of cells and merging into zones is a non-linear procedure, such monotonic growth is not observed for all voids.

We highlight a single void and show how it evolves as axion mass decreases in Fig.~\ref{fig:single_void}. As discussed above, the void increases in size. Further, within the void, there are larger regions devoid of halos, indicating that voids get both larger and emptier of halos. In \S~\ref{sec:voidanalysis}, we will statistically probe both these phenomena through the void-halo correlation function, the void density profile and the void size function. Voids are manifestly not symmetric though an effective radius is approximated from a sphere that has the same volume as the void (indicated by the red circle in Fig.~\ref{fig:single_void}). The void center is defined by the region within the void which is emptiest of halos (see above; center is indicated by the red point in Fig.~\ref{fig:single_void}).

\begin{figure*}
    \centering
    \begin{tabular}{cc}
      \includegraphics[width=0.48\linewidth]{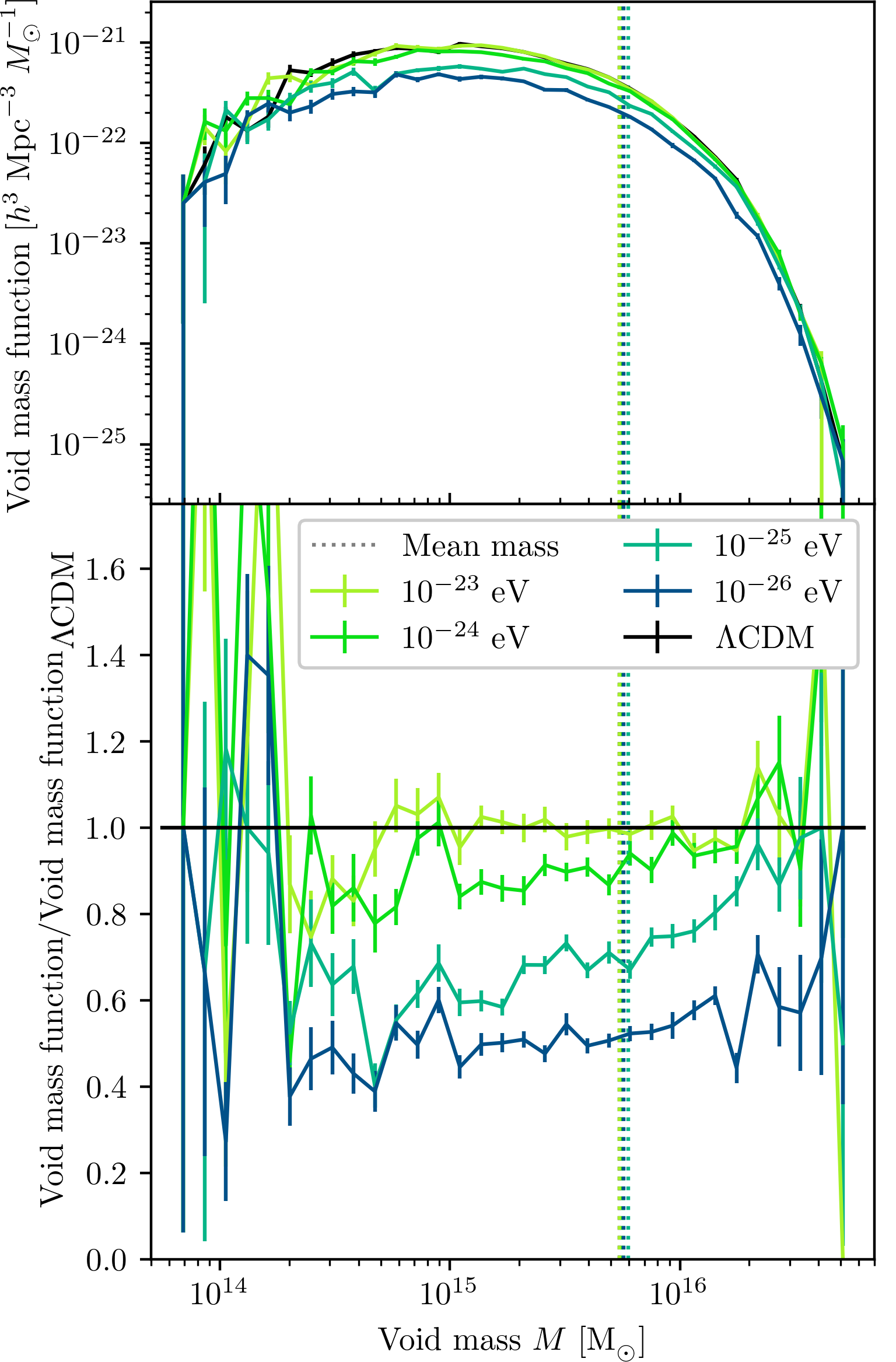}   & \includegraphics[width=0.48\linewidth,trim={0cm 0 0cm 0},clip]{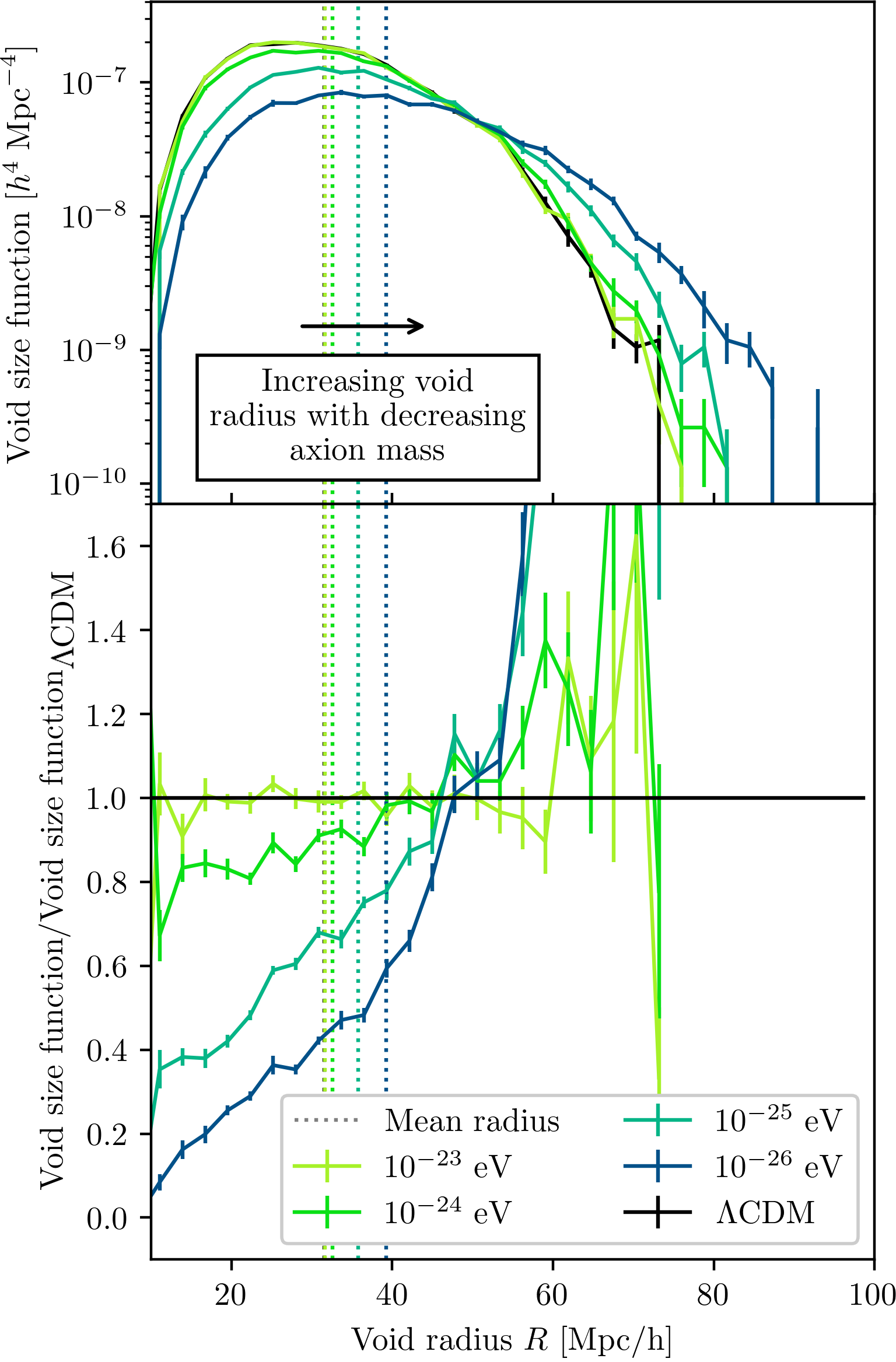} \\
    \end{tabular}
     \caption{\textit{In the left panels}, the void mass function (number density of voids per unit void mass) of simulation \#3 and \#6-9 (see Table~\ref{parameters-table} for simulation parameters), showing the effect of decreasing axion mass. \textit{In the right panels}, we show the void size function (number density of voids per unit void radius) for the same simulations. The \textit{upper panels} show the absolute mass and size functions, while the \textit{bottom panels} show the ratio to the \(\Lambda\)CDM case. We indicate the sample variance by the 68\% confidence limit error bars and the average void mass and radius by dotted lines. While the void size function shifts to larger voids with decreasing axion mass, the void mass function reduces largely in amplitude. Thus, voids grow in size with decreasing axion mass, but they get emptier of halos and fewer in number.\label{fig:VSMF}}
\end{figure*}

\begin{figure}
    \centering
    \includegraphics[width=1\linewidth]{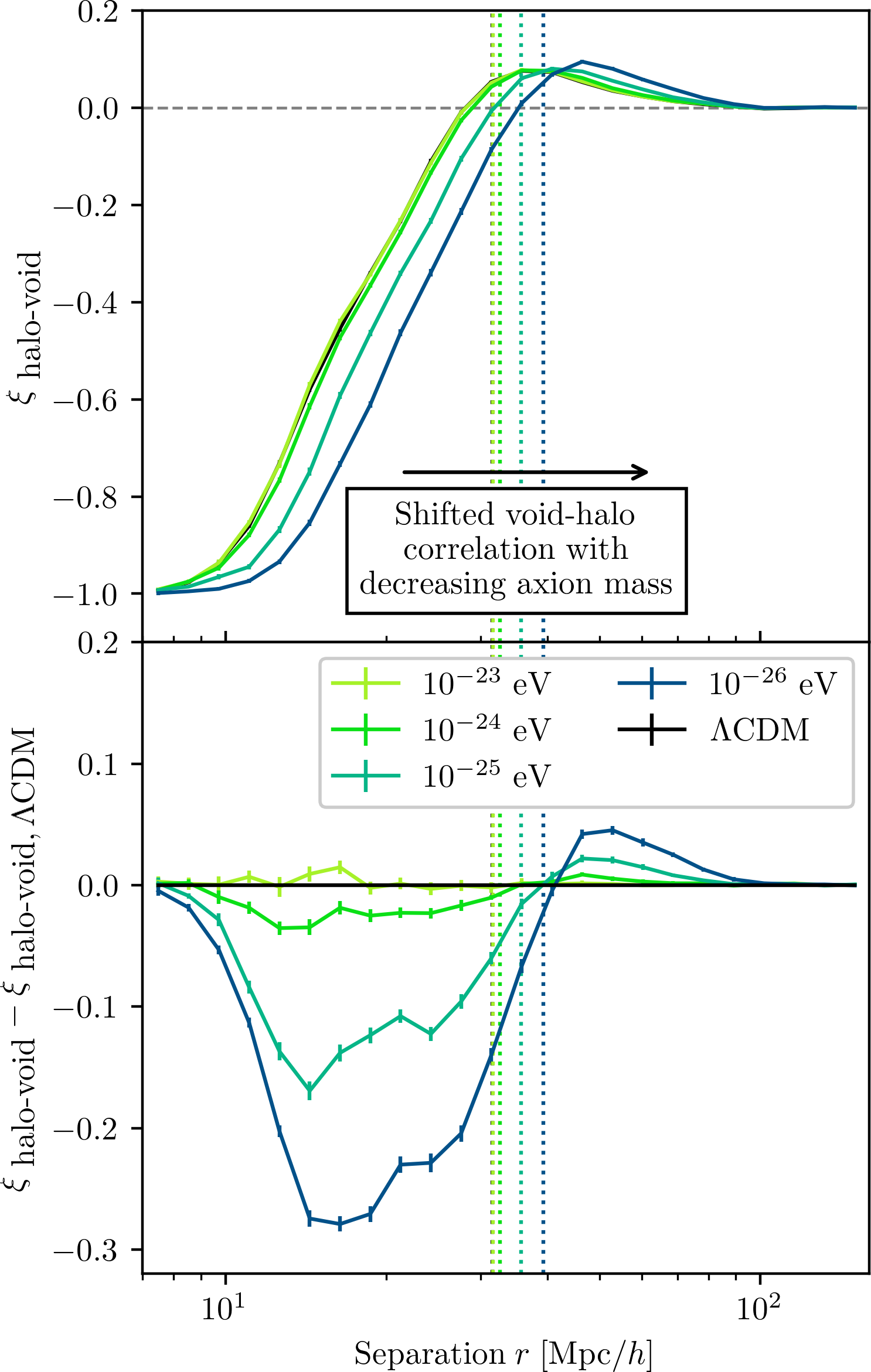}
    \caption{\textit{Top panel}: similar to Fig.~\ref{fig:halohaloCF}, except for the void-halo correlation function. \textit{In the bottom panel}, we show the difference to the \(\Lambda\)CDM case. The dotted lines indicate the average void radius, which sets the scale at which voids and halos become negatively correlated. As the axion mass decreases, the average void radius increases and so the void-halo correlation shifts to larger separation. \label{fig:voidhaloCF}}
\end{figure}

\begin{figure}
    \centering
    \includegraphics[width=1\linewidth]{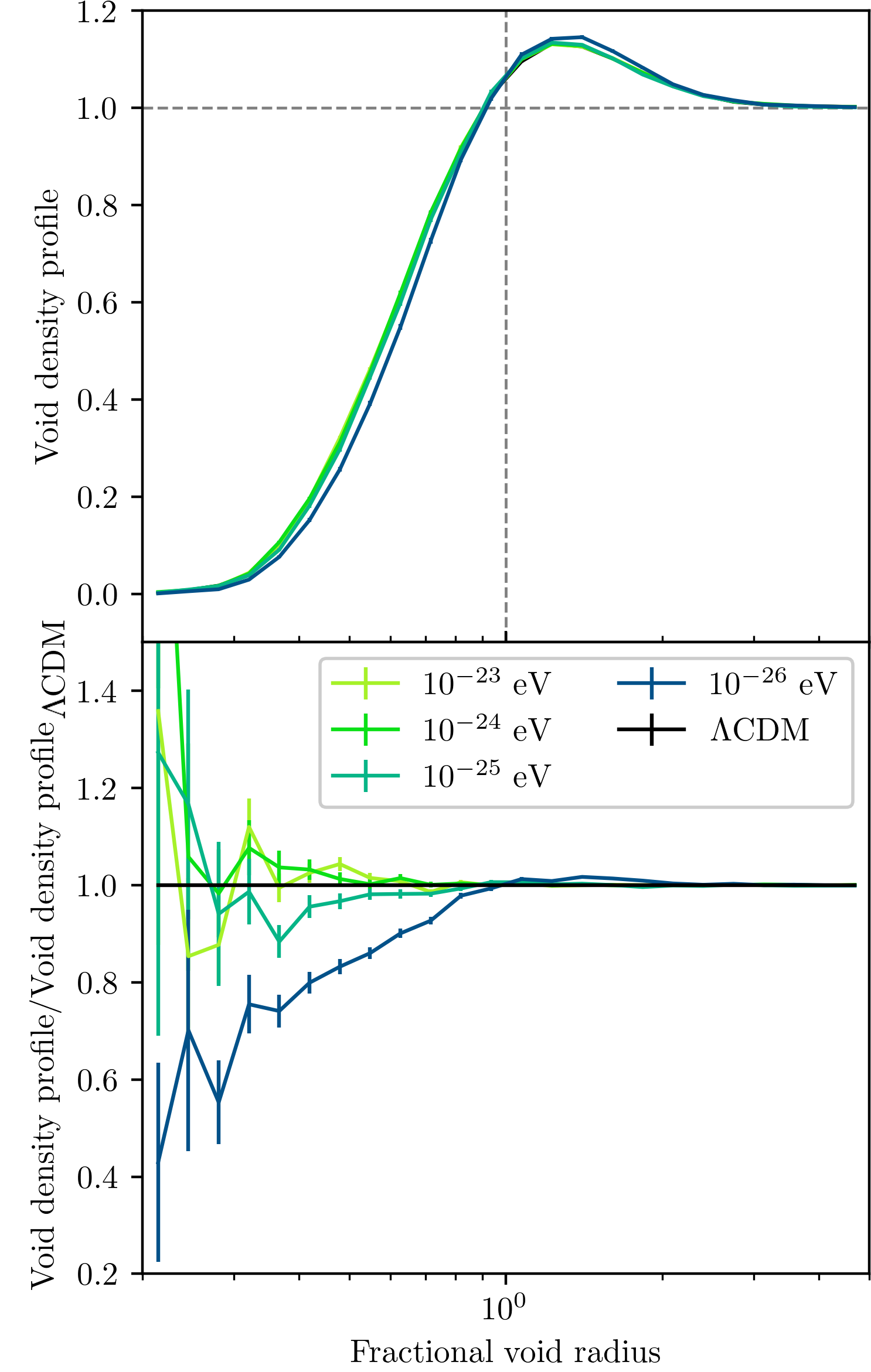}
    \caption{The void density profile (as defined by the local number density of halos in ratio to the global halo number density) of simulation \#3 and \#6-9 (see Table \ref{parameters-table}) as a function of radius from void centers normalized to each void's effective radius, showing the effect of decreasing axion mass. The \textit{upper} panel shows the normalized local halo number density profile averaged over all voids. The \textit{bottom} panel shows the ratio to the \(\Lambda\)CDM case. We indicate the sample variance by the 68\% confidence error bars. As axion mass decreases, the void density profile gets deeper, indicating that the voids get emptier of halos. \label{fig:voidDensityProfile}}
\end{figure}

\begin{figure}
    \centering
    \includegraphics[width=1\linewidth]{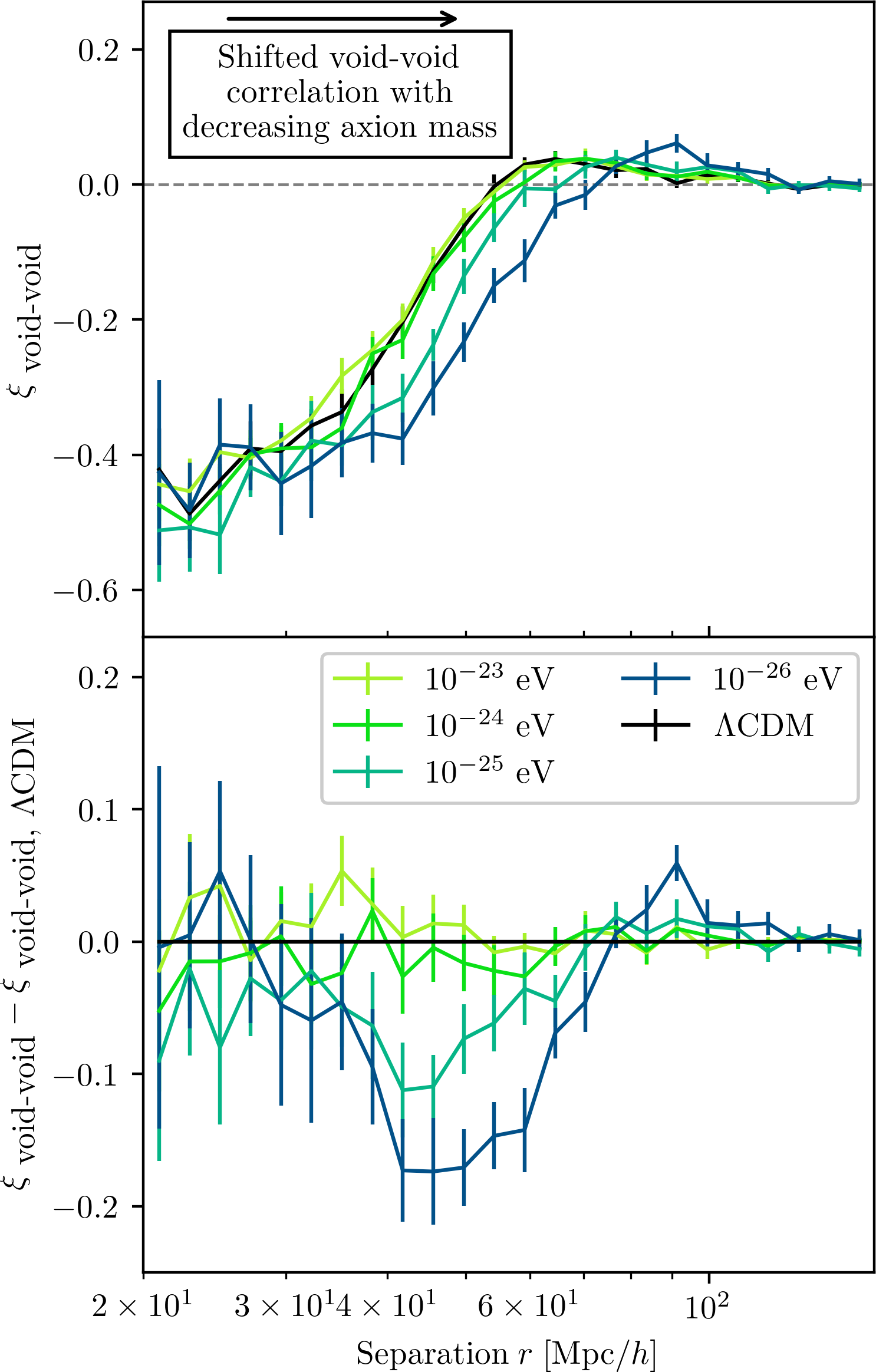}
    \caption{As Figs.~\ref{fig:halohaloCF} and \ref{fig:voidhaloCF}, except for the void-void correlation function. \textit{In the bottom panel}, we show the difference to the \(\Lambda\)CDM case. As axion mass decreases, the average void radius increases and so does the typical separation between void centers. Thus, the void-void correlation shifts to larger separation. \label{fig:voidvoidCF}}
\end{figure}

\section{Void catalog results} \label{sec:voidanalysis}

\subsection{Void mass function and void size function} \label{subsec:voidsize}

In Fig.~\ref{fig:VSMF}, we compare the void mass function and void size function, i.e., the number density of voids respectively as a function of their mass and size, with varying axion mass to the \(\Lambda\)CDM case. As in \S~\ref{sec:halo-corr}, we calculate sample variance by a jackknife estimate (\S~\ref{sec:errors}). As the axion mass decreases, the number of low-mass halos decreases. This reduction in halos causes voids on average to increase in size, driven by the merger of smaller voids, hence the shift to larger radius of the void size function in Fig.~\ref{fig:VSMF}. Indeed, the mean void radius increases from \(31.6\,\mathrm{Mpc}/h\) in the \(\Lambda\)CDM simulation to \(39.2\,\mathrm{Mpc}/h\) in the simulation with \(10^{-26}\,\mathrm{eV}\) axions. In contrast, the mean void mass stays roughly constant as the axion mass changes. The main effect of axions on the void mass function is a reduction in the overall amplitude, indicating fewer voids in total (from the merging of smaller voids), with each on average emptier of halos. In \S~\ref{sec:voidforecast}, we forecast the sensitivity to axion mass from the void size function as measured in large-volume cosmological surveys.

\subsection{Void-halo correlation and void-void correlation} \label{subsec:voidcorr}

We show the correlation function between voids and halos and its dependence on axion mass in Fig.~\ref{fig:voidhaloCF}; and the correlation function between voids and voids and its dependence on axion mass in Fig.~\ref{fig:voidvoidCF}. The void-halo and void-void correlations are calculated using the Landy-Szalay estimator as implemented in the \texttt{corrfunc} code suite (\S~\ref{sec:2pcf}) and we calculate sample variance by a jackknife estimate (\S~\ref{sec:errors}). The void-halo correlation is negative at separations less than about the average void radius because it is less likely than in a random uniform distribution to have halos within a void than outside it. The correlation is maximally negative at small separations since there are no halos forming at the void center by its definition (see \S~\ref{sec:voidfinder}). As the axion mass decreases, the average void radius increases (see Fig.~\ref{fig:VSMF}) and so the void-halo correlation shifts to larger separation. There is a positive correlation bump at and around the average void radius. This bump is consistent with voids being defined by overdense edges (clusters, filaments, walls), where it is more likely than average to have halos. At the largest separations, the positions of halos are uncorrelated with voids.

The void-void correlation is negative at separations less than about \textit{twice} the average void radius because it is less likely than average to have a void center within two average void radii of another void center.\footnote{Voids are defined not to overlap (see \S~\ref{sec:voidfinder}) so this is a statement about the average separation between voids of different sizes.} As the axion mass decreases, the average separation between void centers increases (since the voids themselves are getting bigger) and so the void-void correlation shifts to larger separation. The void-void correlation is much noisier than the void-halo correlation since there are many fewer void pairs in a fixed volume.

In Fig.~\ref{fig:voidDensityProfile}, we normalize the separation of void-halo pairs by each void radius. This normalization removes the effect of increasing average void radius as axion mass decreases. Concretely, for each simulation, we measure the local number density of halos as a function of distance from the center of each void. We then normalize the local halo number density by the global halo number density of the simulation. We then calculate the average of this density profile across all voids in each simulation. We calculate the 68\% confidence limits on this quantity as the standard deviation across all voids. We show this normalized halo number density profile as a function of radius from void centers, i.e. the separation for each halo-void pair. Unlike in Fig.~\ref{fig:voidhaloCF}, we normalize all these separations by the effective radius of the void in the pair. We see that the void density profile, even after normalization to the void effective radii, gets increasingly underdense as axion mass decreases. This trend reflects the fact that voids get emptier of halos as axion mass decreases, which was discussed in \S~\ref{sec:voidfinder} and Fig.~\ref{fig:single_void}. In other words, as axion mass decreases, it is increasingly unlikely to find halos within voids, even after accounting for the fact that voids get bigger.

The correlation functions and density profiles that we measure using halos as biased tracers of the underlying matter field differ from those derived directly from the matter field. \citet{Yang2015WarmthElevating} find that the average void density profile (i.e., the void-matter correlation function for separations less than the average void radius) is shallower in the presence of the scale-dependent matter power spectrum suppression caused by warm dark matter. This effect is due to the matter field being more diffuse than in the CDM case. Since the void density profile is a measure of the amount of matter within a certain distance from the void center, it is equivalent to the void-matter correlation function within the void radius (modulo the normalizations we apply above). Therefore, as with warm dark matter, we expect the void-matter correlation function to be \emph{less negative} and the matter density profiles of voids to be \emph{less underdense} in the presence of ULAs. However, the same trend cannot be expected for biased tracers. In the presence of ULAs, the diffuse matter lying in voids does not coalesce into halos as it does for CDM. Indeed, as discussed in \S~\ref{sec:halo-corr}, ULAs suppress lower-mass halos, which primarily reside in lower-density regions, i.e., voids. Therefore, the number density of halos within voids is lower than in CDM, i.e., the void-halo correlation is \emph{more negative} (Fig.~\ref{fig:voidhaloCF}) and the halo number density profiles of voids are \emph{more underdense} (Fig.~\ref{fig:voidDensityProfile}). The matter field is not one of the outputs of our simulations. We leave the study of the void-matter correlation function with ULAs for future work; although, we note that it is biased tracers that can be observed in surveys.

In \S~\ref{sec:voidforecast}, we model the void size function (\S~\ref{subsec:voidsize}) in order to perform a forecast with cosmological data. We defer to future work the modeling of the void-halo correlation function since this is not directly observable, rather we observe the void-\textit{galaxy} correlation (or correlation with some other biased tracer). Nonetheless, we demonstrate here for the first time the effects of axions on the cross-correlation between a void and a biased tracer (halo) catalog.

\section{Void size function forecast}\label{sec:voidforecast}

\subsection{Void size function model}
\label{sec:VSF_model}

We analytically model the void size function that we measure in the simulations (\S~\ref{subsec:voidsize}) with a modification of the Sheth-Tormen formalism \citep{Sheth1999LargeScale,Sheth2001EllipsoidalCollapse,Sheth2004AHierarchy,Jennings2013TheAbundance,Contarini2022EuclidCosmological}:
\begin{align}
    \frac{\mathrm{d} n}{\mathrm{d}\ln R} = \frac{f(\sigma)}{V(R)}\frac{\mathrm{d} \ln \sigma^{-1}(R_\mathrm{L})}{\mathrm{d} \ln R_{\rm L}}\bigg|_{R_{\rm L}=R_{\rm L}(R)},
    \label{eq:VSF_model}
\end{align}
where \(n\) is the void number density, $R$ is the (nonlinear) void radius, $R_\mathrm{L}$ is the Lagrangian void radius, $f(\sigma)$ is the void multiplicity function and $V(R)$ is the void volume. \(\sigma^2(R_\mathrm{L})\) is the variance of linear matter perturbations on the Lagrangian scale \(R_\mathrm{L}\):
\begin{align}
    \sigma^2(R_\mathrm{L}) = \int \frac{k^2\mathrm{d}k}{2\pi^2} P(k) \left|W(k,R_\mathrm{L})\right|^2, \label{eq:sigma}
\end{align}
where \(k\) is wavenumber, \(P(k)\) is the linear matter power spectrum \citep[evaluated using an interpolator of \texttt{AxionCAMB} presented in][]{Lague:2021frh} and $W(k,R_\mathrm{L})$ is the Fourier transform of a spherical top-hat window function with radius \(R_\mathrm{L}\). By conservation of mass, the Lagrangian and nonlinear void radii are related by the average void mass density $\rho_{\rm v}$ and mean mass density of the Universe $\bar{\rho}$:
\begin{align}
    R_{\rm L}^3 = \frac{\rho_{\rm v}}{\bar{\rho}} R^3.
\end{align}

The void multiplicity function is the distribution of perturbations with variance \(\sigma^2\) that become voids (we have dropped the explicit dependence on $R_\mathrm{L}$ for simplicity). By the excursion set formalism, a perturbation becomes a void at radius \(R_\mathrm{v}\) if it is first crossing the linear void formation underdensity threshold \(\delta^\mathrm{L}_\mathrm{v}\), without having crossed the linear collapse overdensity threshold \(\delta^\mathrm{L}_\mathrm{c}\) at any larger scale. \citet{Sheth2004AHierarchy,Jennings2013TheAbundance} thus calculate \(f(\sigma)\) assuming spherical perturbations in Lagrangian space:
\begin{align}
    f(\sigma) = 2\sum_{i=1}^{\infty} \exp\left(-\frac{(i\pi x)^2}{2}\right)i\pi x^2 \sin\left(i\pi \mathcal{D}\right),
\end{align}
where
\begin{align}
    \mathcal{D} = \frac{|\delta_{\rm v}^{\rm L}|}{\delta_{\rm c}^{\rm L}+|\delta_{\rm v}^{\rm L}|}\,\,\,\mathrm{and}\,\,\,x = \frac{\mathcal{D}}{|\delta_{\rm v}^{\rm L}|}\sigma.
\end{align}
We fix $\delta^{\rm L}_{\rm c} = 1.686$ according to the spherical collapse approximation of a virialized halo in an Einstein-de Sitter cosmology. By a cosmology- and redshift-independent fitting function, the linear void formation threshold is related to the nonlinear void underdensity $\delta^{\rm NL}_{\rm v}$ \citep{Bernardeau1994TheNonlinear}:
\begin{align}
    \delta^{\rm L}_{\rm v} = \mathcal{C}\left[1-\left(1+\delta^{\rm NL}_{\rm v}\right)^{-1/\mathcal{C}}\right], \label{eq:bernardeau-fit}
\end{align}
where $\mathcal{C} \equiv 1.594$.

\begin{figure}
    \centering
    \includegraphics[width=\linewidth]{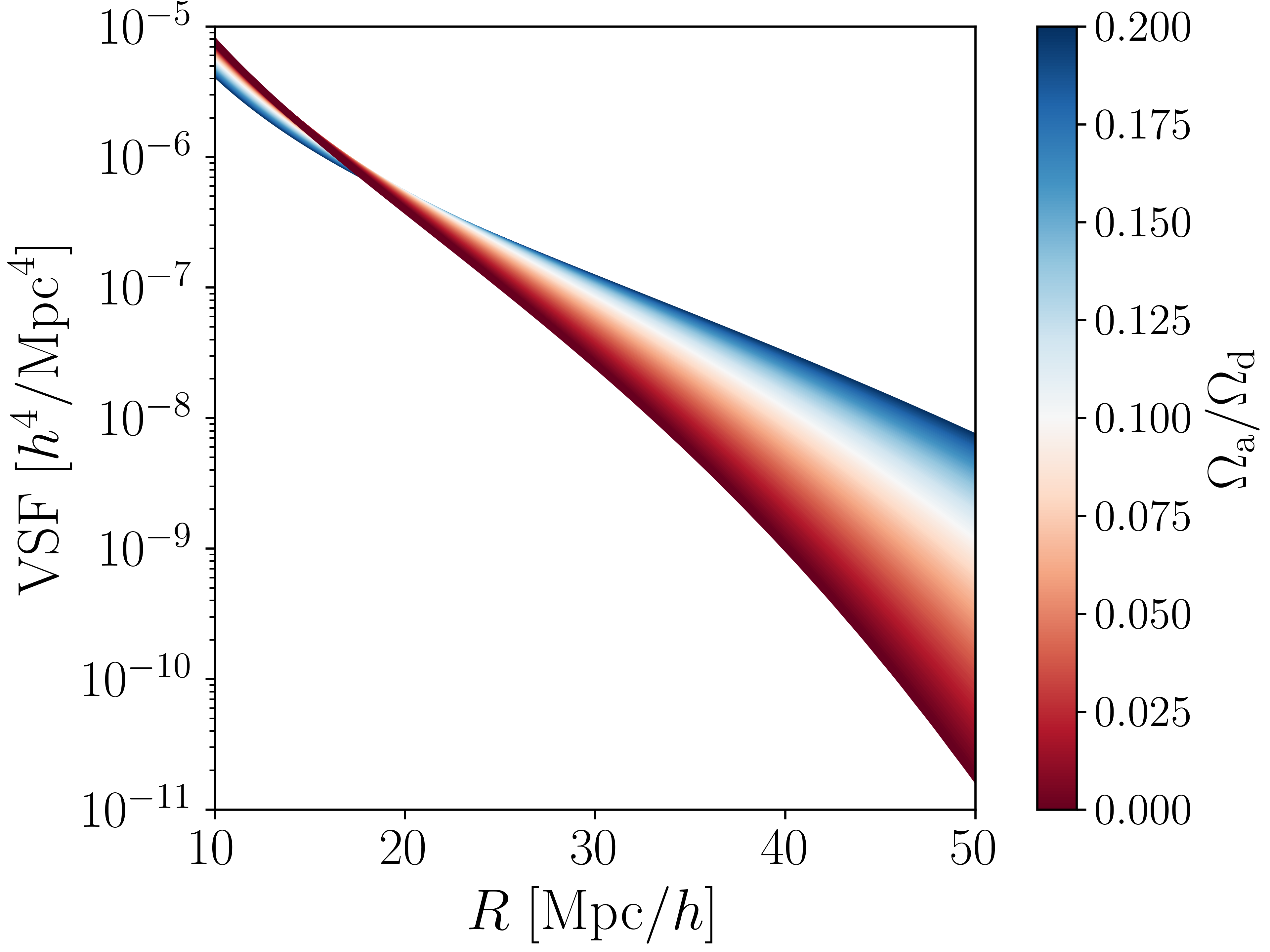}
    \caption{The void size function (VSF) model [Eqs.~\eqref{eq:VSF_model} to \eqref{eq:delta_v-with-b_g}] as a function of void radius \(R\) with varying axion fraction \(\Omega_\mathrm{a} / \Omega_\mathrm{d}\), fixed $m_\mathrm{a} = 10^{-25}\,\mathrm{eV}$, \(B_\mathrm{slope} = 1\), \(B_\mathrm{offset} = 0\), $z=0$ and \(b_\mathrm{h}\) calculated analytically by the \texttt{colossus} cosmology code \citep{Diemer2018COLOSSUS}. For the VSF forecast, we instead fit \(b_\mathrm{h}\) to our simulations. All other cosmological parameters are fixed to the fiducial values given in \S~\ref{sec:simulations} As axion fraction increases, the void size function shifts to larger voids. This result is qualitatively consistent with the simulation results in \S~\ref{subsec:voidsize}, but, as discussed in \S~\ref{sec:VSF_likelihood}, this figure should not be directly compared to Fig.~\ref{fig:VSMF}. This is because the model assumes a certain halo number underdensity in voids that is not \textit{a priori} satisfied in our simulations, which motivates the resizing procedure discussed below.}
    \label{fig:vsf-model-ax-frac}
\end{figure}

We must relate the nonlinear void underdensity expressed as a total matter underdensity $\delta^{\rm NL}_{\rm v}$ to a halo number underdensity $\delta^{\rm NL}_{\rm v,h}$, since we use the positions of halos, not the underlying matter field, to define our voids. To do this, we use the functional form found to reproduce numerical simulation results \citep{2019MNRAS.488.3526C}:
\begin{align}
    \delta^{\rm NL}_{\rm v} = \frac{\delta^{\rm NL}_{\rm v,h} }{B_{\rm slope} b_{\rm h} + B_{\rm offset}},\label{eq:delta_v-with-b_g}
\end{align}
where $b_{\rm h}$ is the halo bias, \(B_{\rm slope}\) and \(B_{\rm offset}\) are parameters that we fit to our simulations in \S~\ref{sec:VSF_likelihood} and we fix $\delta^{\rm NL}_{\rm v,h} = -0.7$ following \citet{Contarini2022EuclidCosmological}. The denominator in Eq.~\eqref{eq:delta_v-with-b_g} is sometimes called the ``punctual'' bias, which is how we will refer to it in this work. The voids in our simulations do not necessarily satisfy $\delta^{\rm NL}_{\rm v,h} = -0.7$ which motivates the resizing procedure as we will discuss in \S~\ref{sec:VSF_likelihood}.

In Fig.~\ref{fig:vsf-model-ax-frac}, we show the void size function model (introduced in Eqs.~\eqref{eq:VSF_model} to \eqref{eq:delta_v-with-b_g}) with varying axion fraction and fixed \(m_\mathrm{a} = 10^{-25}\,\mathrm{eV}\). ULA (and other cosmological) parameters affect the void size function model in two ways. First, the linear matter power spectrum suppression from axions reduces the variance of fluctuations \(\sigma^2\) on small scales (see Eq.~\eqref{eq:sigma}). Second, axions increase the halo bias \(b_\mathrm{h}\) (see \S~\ref{sec:axion_halo_bias}). The combined effect is that increasing axion fraction tends to increase the typical void size and thus shift the void size function to larger radii. This result is qualitatively similar to the effect of axion mass demonstrated by the simulation results in \S~\ref{subsec:voidsize}. However, we defer a detailed comparison between simulations and theory to \S~\ref{sec:VSF_likelihood}.

\subsection{Void size function likelihood and prior}
\label{sec:VSF_likelihood}

In order to compare our simulations to the theory introduced in \S~\ref{sec:VSF_model}, we resize our voids so that the void underdensity condition \(\delta_\mathrm{v,h}^\mathrm{NL} = -0.7\) applies to all voids, following \cite{Ronconi2017CosmologicalExploitation}. We first remove spurious voids that we identify as having a size smaller than twice the mean inter-halo separation (\(\sim 15\,\mathrm{Mpc} / h\)). We then rescale each void radius so that the mean halo number underdensity of each void is $\delta_\mathrm{v,h}^\mathrm{NL}=-0.7$. The halo number density within each void is found by counting the number of halos within a sphere around the void center. The degree to which each void is resized varies, but typically voids are reduced in size in order to make them more underdense, which is consistent with \cite{Ronconi2017CosmologicalExploitation}. As discussed in \S~\ref{subsec:voidcorr}, the centers of voids are more underdense and so reducing the effective radius of a void picks this region out. Finally, we remove overlapping voids by keeping the larger one in the catalog \citep[this choice does not significantly alter the final VSF,][]{Ronconi2017CosmologicalExploitation}. After resizing the void catalog, we bin the void size function in \(N_\mathrm{bins}\) equally-spaced void radius bins from \(25\,\mathrm{Mpc}/h\) to \(40\,\mathrm{Mpc}/h\). As mock data for a void size function forecast, we thus use measurements of the (resized) void size function from two mass-peak patch simulations (\S~\ref{subsec:voidsize}): the $\Lambda$CDM reference cosmology (simulation \#3; Table \ref{parameters-table}) and the simulation with $m_\mathrm{a}=10^{-25}\,\mathrm{eV}$ and \(\Omega_\mathrm{a}/\Omega_\mathrm{d} = 0.1\) (simulation \#6).

We compare the mock observed void number (in a void radius bin $N_{\rm data}(R_i)$) with the model expectation (see \S~\ref{sec:VSF_model}; $N_{\rm model}(R_i, \boldsymbol{\theta})$) assuming a Poisson likelihood with independent data bins \citep{Sahlen2016ClusterVoid}:
\begin{align}
    \mathcal{L}(N_{\rm data}|\boldsymbol{\theta}) = \prod_{i=1}^{N_{\rm bins}} \frac{N_{\rm model}(R_i, \boldsymbol{\theta})^{N_{\rm data}(R_i)} e^{-N_{\rm model}(R_i, \boldsymbol{\theta})}}{N_{\rm data}(R_i)!},
    \label{eq:likelihood}
\end{align}
where \(R_i\) is the \(i^\mathrm{th}\) void radius bin, $\boldsymbol{\theta}$ is a vector of cosmological and ULA parameters and the total number of voids \(N\) is the integral over the volume of the observational survey (see below) of the void number density (see Fig.~\ref{fig:VSMF} and Eq.~\eqref{eq:VSF_model}). We vary the cosmological parameters $\Omega_{\rm m}$ and $\ln 10^{10} A_\mathrm{s}$ (\(A_\mathrm{s}\) is the primordial power spectrum amplitude) and sometimes the axion fraction $\Omega_{\rm a}/\Omega_\mathrm{d}$ (always fixing \(m_\mathrm{a} = 10^{-25}\,\mathrm{eV}\)). All other cosmological parameters are fixed to their fiducial values since they do not strongly affect the void size function, following \citet{Contarini2022EuclidCosmological}. We sometimes derive the linear matter power spectrum amplitude today \(\sigma_8\).

We fix the bias parameters \(b_\mathrm{h}, B_\mathrm{slope}, B_\mathrm{offset}\) (see Eq.~\eqref{eq:delta_v-with-b_g}). It is feasible to marginalize these parameters with a simulation suite (or data) at multiple redshifts. \citet{Contarini2022EuclidCosmological} derive a prior on the punctual bias parameters from their multi-redshift simulations. Since we have simulations only at $z=0$, we instead fit the punctual bias parameters directly to the simulations. The halo bias \(b_\mathrm{h}\) is fit to the halo-halo power spectrum as discussed in \S~\ref{sec:axion_halo_bias}. The void punctual bias parameters \(B_\mathrm{slope}, B_\mathrm{offset}\) are fit to the mock observed void size function using the above likelihood with cosmological parameters fixed to their true input and \(b_\mathrm{h}\) fixed to its value from \S~\ref{sec:axion_halo_bias}. When fitting to the mock \(\Lambda\)CDM data, we find \(B_\mathrm{slope} = 0.520\) and \(B_\mathrm{offset} = 0.925\), giving a punctual bias consistent with the \textit{Euclid} Flagship simulations \citep{Contarini2022EuclidCosmological}. We use the same fit of the punctual bias values when inferring cosmological parameters from the mock axion data. Ignoring the axion impact on the punctual bias fit is incorrect but reflects what will occur in future data analyses \citep{Contarini2022EuclidCosmological}. In other words, we evaluate the posterior distribution of cosmological and ULA parameters at a maximum likelihood point in void size function bias parameters (using halo-halo correlations when determining \(b_\mathrm{h}\)).

When varying axion fraction in the cosmological model, we approximate the use of CMB data with the prior $\Omega_\mathrm{m} \sim \mathcal{N}(0.31, 0.002)$ and $\ln 10^{10} A_\mathrm{s} \sim \mathcal{N}(3.043, 0.3)$, following forecast \(\Lambda\)CDM constraints from the Simons Observatory \citep{Ade2019SimonsObs}. Doing this breaks a strong degeneracy between $\Omega_{\rm m}$ and \(\Omega_\mathrm{a}/\Omega_\mathrm{d}\). We otherwise use a uniform prior on \(\Omega_\mathrm{a}/\Omega_\mathrm{d}\). We sample the posterior distribution using the Markov chain Monte Carlo algorithm \texttt{Zeus} \citep{karamanis2020ensemble,karamanis2021zeus}. We evaluate sampling convergence by checking that each chain contains at least 50 independent autocorrelation lengths.

\begin{table}
    \centering
    \begin{tabular}{|c|c|c|c|}
       \hline
       Survey & $z$ range & $V_{\rm eff}$ \(\left(\mathrm{Gpc}^{3}\right)\) & $\sqrt{V_{\rm eff}/V_{\rm sim}}$ \\
       \hline 
       eBOSS-like & $0.4 - 1.1$ & $8$ & $ 0.97$ \\
       DESI-like  & $0.1 - 2.1$ & $42$ & $ 2.2$\\
       \textit{Euclid}-like  & $0.9 - 1.7$ & $73$ & $ 2.9$ \\
       \hline
    \end{tabular}
    \caption{The approximate observational settings that we consider with (\textit{from left to right}) the true redshift range of the survey, the effective volume of each survey \(V_\mathrm{eff}\) and the square root of the ratio between the effective survey volume and the volume of the baseline mass-peak patch simulations \#3 and \#6 (see Table \ref{parameters-table}), which is equivalent to the ratio of data uncertainties between simulation and survey (see the main text for discussion). \textit{From top to bottom}, we consider three different survey settings.}
    \label{tab:surveys-volume}
\end{table}

We consider approximations to three completed and ongoing surveys for the forecast. The first setting is the Baryon Oscillation Spectroscopic Survey \citep[BOSS;][]{2013AJ....145...10D,Alam2017BOSS} and its extension eBOSS \citep{Dawson2016eBOSS}; the second setting is the Dark Energy Spectroscopic Instrument \citep[DESI;][]{Aghamousa2016DESI}; and the third setting is \textit{Euclid} \citep{euclid}. For an eBOSS-like survey, we use the effective volume of the BOSS galaxies and the ELG and LRG galaxy samples of eBOSS \citep{Alam2021CompletedSDSS-IV}. For a DESI-like survey, we use the effective volume of the full BGS, ELG, LRG and QSO samples \citep{AbdulKarim2025DESIDR2}. For a \textit{Euclid}-like survey, we use the effective volume calculation of \citet{Chudaykin2019MeasuringNeutrino} and limit the galaxy redshift range to match more closely the \textit{Euclid} void size function forecast of \citet{Contarini2022EuclidCosmological}. The redshift range and effective volume \(V_\mathrm{eff}\) of each survey setting are given in Table \ref{tab:surveys-volume}.

\begin{figure}
    \centering
    \includegraphics[width=\linewidth]{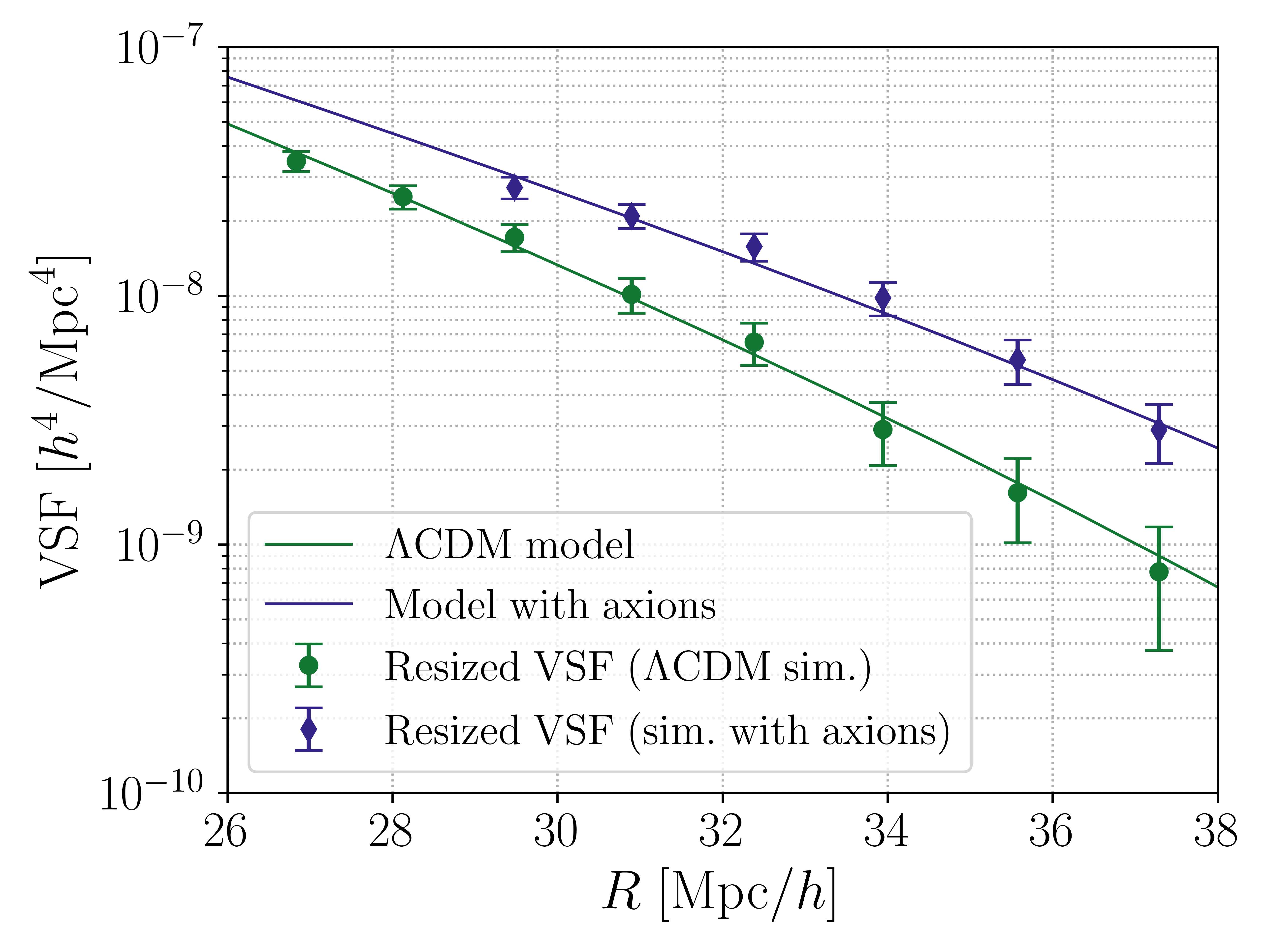}
    \caption{Mock void size function for an eBOSS-like survey (VSF; points with 68\% confidence limit Poisson error bars; see the main text for details) as a function of void radius \(R\), given \(\Lambda\)CDM (\textit{green}) and axion (\(m_\mathrm{a} = 10^{-25}\,\mathrm{eV}\); \(\Omega_\mathrm{a}/\Omega_\mathrm{d} = 0.1\); \textit{blue}) simulations (after resizing, see \S~\ref{sec:VSF_likelihood}), compared to the best-fit model (shown as lines and described in \S~\ref{sec:VSF_model}).}
    \label{fig:vsf-with-data}
\end{figure}

To approximate each observational setting, we multiply $N_{\rm data}$ and $N_{\rm model}$ for the baseline \(\Lambda\)CDM and axion simulations by the ratio between each survey effective volume and the simulation volume (Table \ref{tab:surveys-volume}). We thus assume that all voids are found at $z=0$, ignoring redshift evolution in each survey. In this approximate first forecast, we also ignore all systematic uncertainties, thus assuming a complete void catalog from the largest void that we consider (\(R=40\,h^{-1}\,\mathrm{Mpc}\)) to the smallest void that we consider (\(R=25\,h^{-1}\,\mathrm{Mpc}\) for \(\Lambda\)CDM; \(R=28\,h^{-1}\,\mathrm{Mpc}\) for the axion simulation; the axion simulation is slightly less well converged below this size, see Appendix \ref{app:simulation}). These data cuts are similar to those in \citet{Contarini2022EuclidCosmological}. Most importantly, we ignore the galaxy-halo connection that must be modeled for comparison with real data. We discuss in more detail the approximations made in this forecast and their implications in \S~\ref{sec:approx_discussion}.

Fig.~\ref{fig:vsf-with-data} shows the mock BOSS + eBOSS void size function compared to the best-fit model (\S~\ref{sec:VSF_model}) for both the \(\Lambda\)CDM and axion cases. There is broad agreement between mock data and theory; we will discuss the robustness of the theoretical model further in \S~\ref{sec:forecast_results}. The fractional uncertainty on the number of voids \(N_\mathrm{data}\) observed in a given void radius bin is \(\frac{1}{\sqrt{N_\mathrm{data}}} = \frac{1}{\sqrt{n V_\mathrm{eff}}}\), where \(n\) is the number density of voids. We use this calculation to generate the error bars in Fig.~\ref{fig:vsf-with-data}, though we note that we use the full Poisson likelihood in Eq.~\eqref{eq:likelihood} for the forecast. We give in Table \ref{tab:surveys-volume} the ratio \(\sqrt{\frac{V_\mathrm{eff}}{V_\mathrm{sim}}}\) (the baseline simulation volume \(V_\mathrm{sim} = \left(2048\,\mathrm{Mpc}\right)^3\)) and thus the ratio of the data uncertainties between simulation and survey.

\subsection{Forecast results}
\label{sec:forecast_results}

\begin{figure*}
    \centering
    \subfigure{
    \includegraphics[width=0.49\linewidth]{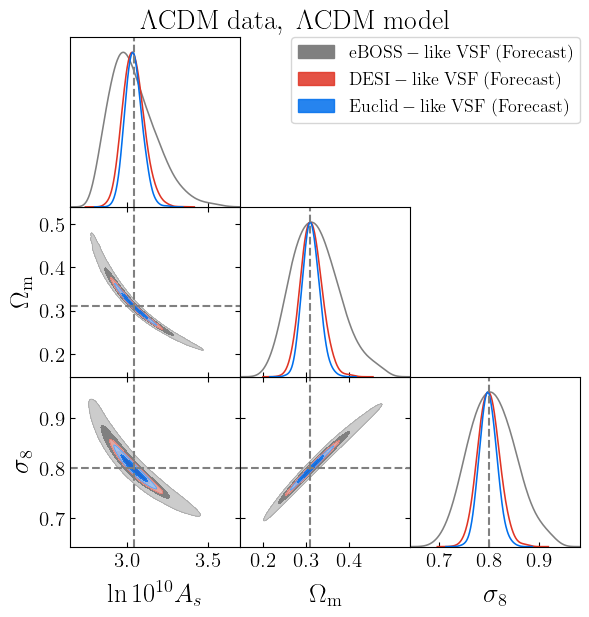}}
    \subfigure{
    \includegraphics[width=0.49\linewidth]{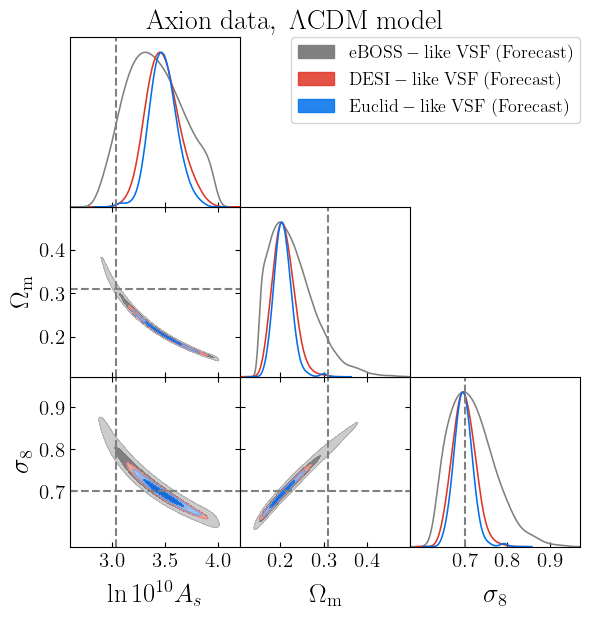}}
    \subfigure{
    \includegraphics[width=0.49\linewidth]{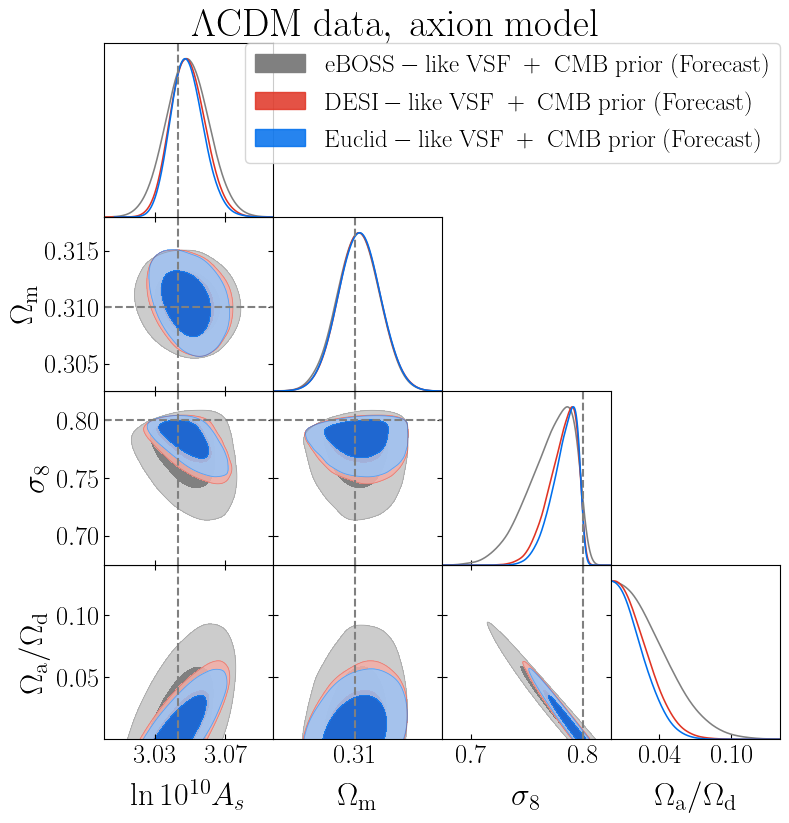}} 
    \subfigure{
    \includegraphics[width=0.49\linewidth]{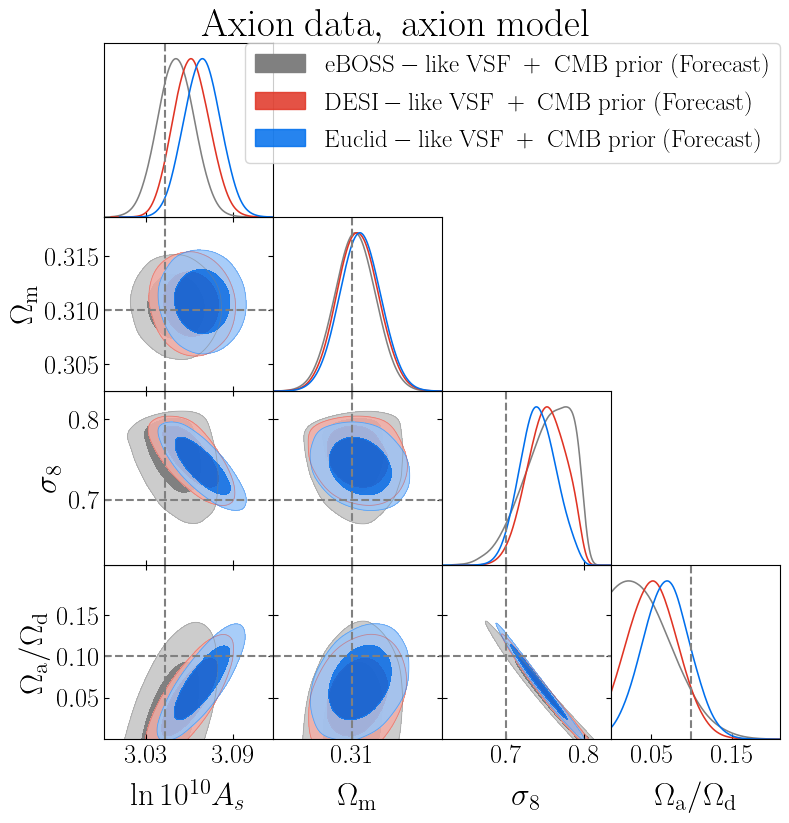}}
        \caption{Forecast marginalized posterior distributions of \(\Lambda\)CDM and axion parameters (always fixing \(m_\mathrm{a} = 10^{-25}\,\mathrm{eV}\)) given the void size function measured from eBOSS-like (\textit{grey}), DESI-like (\textit{red}) and \textit{Euclid}-like (\textit{blue}) galaxy surveys (Table \ref{tab:surveys-volume}). We consider two analysis settings each for the mock data and the model used for inference: the mock data are $\Lambda$CDM-like (\textit{left column}; simulation \#3) or axion-like (\textit{right column}; simulation \#6); the model inferred is \(\Lambda\)CDM (\textit{top row}) or an axion cosmology (\textit{bottom row}). The darker and lighter shaded areas respectively indicate the 68\% and 95\% credible regions. The black dotted lines indicate the true parameters of the mock data. For the axion model analyses, to break parameter degeneracy given the void size function alone, we use a CMB prior on \(\Lambda\)CDM parameters given upcoming Simons Observatory data (\S~\ref{sec:VSF_likelihood}). \textit{From left to right, top to bottom}, we thus forecast: (i) the \(\Lambda\)CDM constraining power; (ii) the bias if axions exist but are not modeled (\(\Lambda\)CDM assumed); (iii) the axion density limit if they do not exist; and (iv) the axion detection constraining power if they do exist and are included in the model.
    \label{fig:validation-mcmc}}
\end{figure*}

\begin{table}
    \centering
    \begin{tabular}{|c|c|c|c|}
        \hline
         Mock data & Survey & $\Omega_\mathrm{a} / \Omega_\mathrm{d}$ $[1 \sigma]$ & $\Omega_\mathrm{a} / \Omega_\mathrm{d}$ $[2 \sigma]$\\ \hline
         \(\Lambda\)CDM & eBOSS-like & $ < 0.039$ & $ < 0.075$\\
         \(\Lambda\)CDM &  DESI-like & $ < 0.026$ & $ < 0.051$\\
         \(\Lambda\)CDM &  \textit{Euclid}-like & $ < 0.023$ & $ < 0.046$\\ \hline
         \(m_\mathrm{a} = 10^{-25}\,\mathrm{eV}\) & eBOSS-like & $ < 0.062$ & $ < 0.114$\\
         \(m_\mathrm{a} = 10^{-25}\,\mathrm{eV}\) & DESI-like & $  0.055^{+0.026}_{-0.034}$ & $ < 0.106$ \\
         \(m_\mathrm{a} = 10^{-25}\,\mathrm{eV}\) & \textit{Euclid}-like & $  0.069 \pm 0.029$ & $  0.069^{+0.055}_{-0.057}$\\
         \hline
    \end{tabular}
    \caption{Forecast \(1 \sigma\) and \(2 \sigma\) constraints on the axion fraction \(\Omega_\mathrm{a} / \Omega_\mathrm{d}\) if axions do not exist (\(\Lambda\)CDM; \textit{top rows}; analysis (iii) in Fig.~\ref{fig:validation-mcmc}) and or if they do exist (\(m_\mathrm{a} = 10^{-25}\,\mathrm{eV}\), \(\Omega_\mathrm{a} / \Omega_\mathrm{d} = 0.1\); \textit{bottom rows}; analysis (iv) in Fig.~\ref{fig:validation-mcmc}). We consider the three galaxy survey settings in Table \ref{tab:surveys-volume} using a CMB prior on \(\Lambda\)CDM parameters given upcoming Simons Observatory data (\S~\ref{sec:VSF_likelihood}). We find that the combination of a \textit{Euclid}-like void size function measurement and \(\Lambda\)CDM constraints from Simons Observatory can improve axion fraction limits by about a factor of two over the current strongest bounds \citep{Winch:2024mrt} and that a \(2 \sigma\) preference is forecast for a 10 \% contribution of axions with \(m_\mathrm{a} = 10^{-25}\,\mathrm{eV}\).}
    \label{tab:fax-bounds}
\end{table}

We evaluate the capacity of the three surveys described in Table~\ref{tab:surveys-volume} to constrain \(\Lambda\)CDM and axion parameters. Fig.~\ref{fig:validation-mcmc} thus shows forecast marginalized posterior distributions for four analysis settings: (i) using a \(\Lambda\)CDM model where the mock data are \(\Lambda\)CDM-like -- simulation \#3; (ii) using a \(\Lambda\)CDM model where the mock data are axion-like -- simulation \#6; (iii) using an axion model where the mock data are \(\Lambda\)CDM-like -- also simulation \#3; and (iv) using an axion model where the mock data are axion-like -- also simulation \#6. We summarize forecast axion constraints in Table~\ref{tab:fax-bounds}.

For the \(\Lambda\)CDM analysis presented in (i), we find that the recovered posteriors are consistent with the input truth, indicating an unbiased inference. There is a strong positive degeneracy between \(\sigma_8\) and \(\Omega_\mathrm{m}\) \citep[consistent with][]{Contarini2022EuclidCosmological} and a strong negative degeneracy between \(\sigma_8\) and \(A_s\) as expected (i.e., boosting the linear matter power spectrum today is compensated by reducing the amplitude of the primordial power spectrum). As the effective survey volume increases from eBOSS to DESI to \textit{Euclid}, the number of observed voids increases and so the parameter constraints improve with no change in the degeneracy. For the analysis presented in (ii), where we inject the effects of axions (\(m_\mathrm{a} = 10^{-25}\,\mathrm{eV}\); \(\Omega_\mathrm{a} / \Omega_\mathrm{d} = 0.1\)) into the mock data but not the assumed model, we find a bias in the inference as expected. Axions suppress the value of \(\sigma_8\) through their effects on the linear matter power spectrum \citep[see \S~\ref{sec:halo-corr} and][]{Rogers:2023ezo}. The lower value of \(\sigma_8\) is correctly recovered by all three survey settings. However, this comes at the expense of overestimating \(A_s\) and underestimating \(\Omega_\mathrm{m}\), moving along the \(\Lambda\)CDM parameter degeneracy directions.

For the analysis presented in (iii), where axions are included in the assumed model, but no axions are injected into the mock data, the Simons Observatory \(\Lambda\)CDM prior breaks the \(\Lambda\)CDM parameter degeneracies present with void size function data alone. These degeneracies are otherwise exacerbated by the additional degeneracy with the axion fraction. We observe the usual negative degeneracy between axion fraction and \(\sigma_8\), where increasing the amount of axions suppresses the linear matter power spectrum and thus lowers \(\sigma_8\). We report the forecast bounds on axion fraction in this case in Table \ref{tab:fax-bounds}, finding the strongest limit comes from \textit{Euclid}, the experiment with the largest survey volume. The forecast \(2 \sigma\) limit is about a factor of two stronger than the current best limit from the combination of \textit{Planck} CMB and Hubble Space Telescope galaxy ultraviolet (UV) luminosity function data \citep{Winch:2024mrt}.

For the analysis presented in (iv), where axions are injected into the mock data and are included in the model, we forecast the ability to detect axions with \(m_\mathrm{a} = 10^{-25}\,\mathrm{eV}\) that constitute 10 \% of the dark matter. We see the same degeneracies as in analysis (iii). A DESI-like survey with a CMB \(\Lambda\)CDM prior can recover the axion signature with \(\sim 1 \sigma\) preference, while a \textit{Euclid}-like survey with the same prior can recover the signature with \(\sim 2 \sigma\) preference. There is however a slight bias in the inference of \(A_s\). We attribute this slight bias to the void punctual bias fit discussed in \S~\ref{sec:VSF_likelihood}. In order to mimic the planned data analysis \citep{Contarini2022EuclidCosmological}, we use the best fit punctual bias from the \(\Lambda\)CDM simulation for the axion analysis. Our results suggest that this is sub-optimal given the precision of \textit{Euclid} data and we instead advocate that the punctual bias should be fully marginalized for extended cosmology analyses.

\begin{figure}
    \centering
    \includegraphics[width=1.0\linewidth]{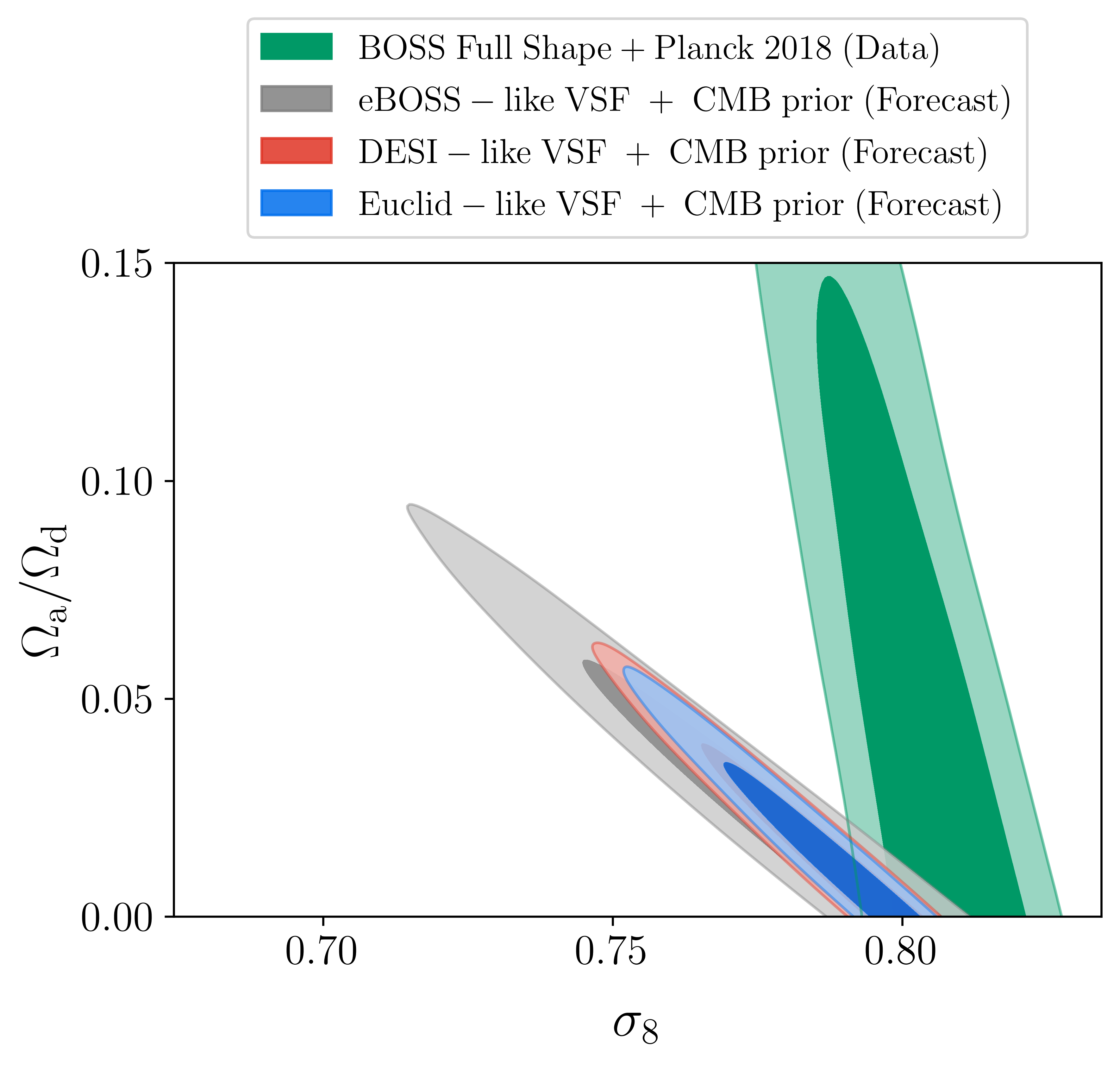}
    \caption{Comparison between the strongest galaxy survey data constraints on the axion fraction \(\Omega_\mathrm{a} / \Omega_\mathrm{d}\) (\textit{green}) and the void size function (VSF) forecasts in this work (eBOSS-like + Simons Observatory (SO) prior: \textit{grey}; DESI-like + SO prior: \textit{red}; \textit{Euclid}-like + SO prior: \textit{blue}; see \S~\ref{sec:forecast_results} for discussion; we show forecast analysis (iii), i.e., the axion limit if they do not exist). We always fix \(m_\mathrm{a} = 10^{-25}\,\mathrm{eV}\). The galaxy constraints are presented in \citet{Rogers:2023ezo}, using a combination of \textit{Planck} CMB and BOSS full-shape galaxy power spectrum and bispectrum multipole data. The darker and lighter shaded areas respectively indicate the 68\% and 95\% credible regions of the marginalized posterior distributions of \(\Lambda\)CDM and axion parameters. We forecast improvements in axion limits with VSF data and a different degeneracy between \(\Lambda\)CDM and axion parameters indicating that future combined analyses will be most powerful. \label{fig:combination-with-cmb}}
\end{figure}

\section{Discussion} \label{sec:discussion}

\subsection{Comparison of voids with other large-scale structure probes}

In Fig.~\ref{fig:combination-with-cmb}, we compare the forecast void size function constraints on \(\Lambda\)CDM and axion parameters (see the analysis presented as (iii) in \S~\ref{sec:forecast_results}) to the current strongest axion limits from galaxy survey data at \(m_\mathrm{a} = 10^{-25}\,\mathrm{eV}\). These limits come from a combination of \textit{Planck} CMB temperature, polarization and lensing angular power spectra and BOSS full-shape galaxy power spectrum and bispectrum multipoles \citep{Rogers:2023ezo}. We forecast that the VSF, even with existing eBOSS data, can significantly improve axion fraction limits from galaxy surveys. This improvement is partly driven by the \textit{linearity of voids} relative to galaxies/overdensities. We model the VSF with excursion-set theory and linear halo and void punctual biases. In contrast, \citet{Rogers:2023ezo} model the galaxy power spectrum and bispectrum using the mildly non-linear effective field theory of large-scale structure \citep{Baumann:2010tm,Ivanov:2022mrd}, which introduces fifty-two bias, counterterm and noise nuisance parameters that are marginalized over with priors derived from numerical simulations. This different modeling approach is needed as galaxies trace overdense, more non-linearly evolved parts of the cosmic web, while voids trace underdense parts less affected by non-linearities at any given scale. Further, Fig.~\ref{fig:combination-with-cmb} illustrates that, since galaxies and voids trace different parts of the large-scale structure, they contain complementary information about the nature of dark matter and thus the degeneracy direction between \(\Lambda\)CDM and axion parameters is different in each case. This complementarity suggests that a future combination of galaxy and void data will be most powerful. We leave to future work a quantitative study of such a joint analysis.

\begin{figure}
    \centering
    \includegraphics[width=\linewidth]{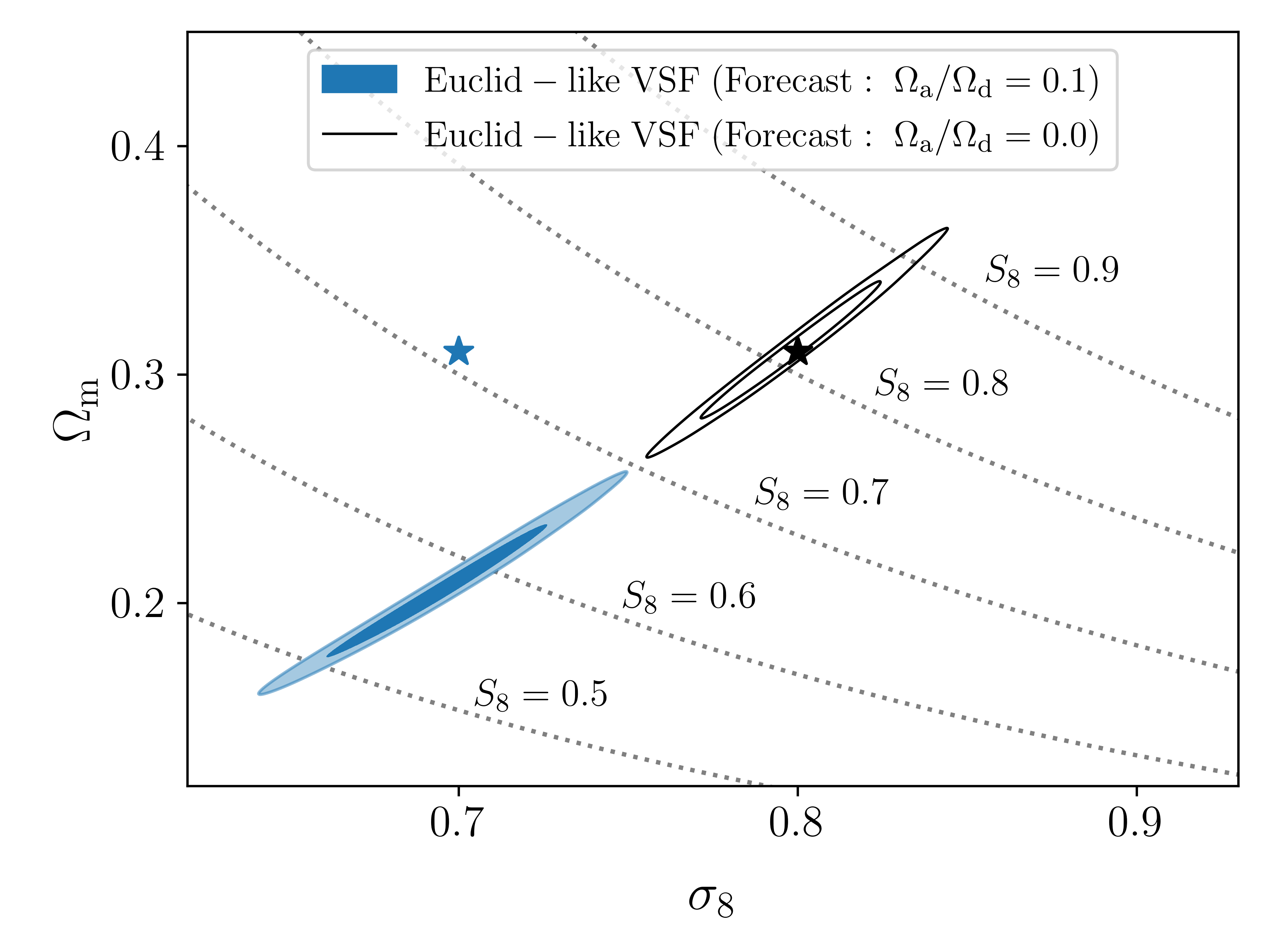}
    \caption{Illustration of the axion-induced bias in parameter inference caused by assuming a $\Lambda$CDM model if axions exist (effect exaggerated given current observational limits). The \textit{black} contours show the forecast posterior distribution of \(\sigma_8\) and \(\Omega_\mathrm{m}\) given \textit{Euclid}-like \(\Lambda\)CDM VSF data and assuming a \(\Lambda\)CDM model (analysis (i) in \S~\ref{sec:forecast_results}; \(\Omega_\mathrm{a} / \Omega_\mathrm{d} = 0\)). The \textit{blue} contours show the forecast posterior given \textit{Euclid}-like axion VSF data and assuming a \(\Lambda\)CDM model (analysis (ii); \(m_\mathrm{a} = 10^{-25}\,\mathrm{eV}\); \(\Omega_\mathrm{a} / \Omega_\mathrm{d} = 0.1\)). The inner and outer contours respectively indicate the 68\% and 95\% credible regions of the marginalized posterior and the stars indicate the true parameters of the mock data. The dotted lines indicate contours of constant \(S_8 \equiv \sigma_8 \left(\frac{\Omega_\mathrm{m}}{0.3}\right)^\frac{1}{2}\), which is (approximately) the parameter combination most precisely constrained by galaxy weak lensing experiments. The VSF posteriors are approximately orthogonal to those from weak lensing indicating that future combined analyses will be most powerful.
    }
    \label{fig:combination-with-WL}
\end{figure}

In Fig.~\ref{fig:combination-with-WL}, we compare the forecast \textit{Euclid}-like void size function \(\Lambda\)CDM constraints if axions (\(m_\mathrm{a} = 10^{-25}\,\mathrm{eV}\)) do (\textit{blue} contours; \(\Omega_\mathrm{a} / \Omega_\mathrm{d} = 0.1\)) or do not (\textit{black}; \(\Omega_\mathrm{a} / \Omega_\mathrm{d} = 0\)) exist (the analyses presented as (i) and (ii) in \S~\ref{sec:forecast_results}). As discussed in \S~\ref{sec:forecast_results}, the presence of axions lowers the true value of \(\sigma_8\), which is correctly inferred. However, since axions are not included in the model, the contour moves along the VSF \(\Lambda\)CDM parameter degeneracy direction and thus comes at the expense of underestimating \(\Omega_\mathrm{m}\). \citet{Rogers:2023ezo} point out that lower values of \(S_8\) inferred from large-scale structure data can be a signature of ultra-light axions. These VSF forecasts are consistent with these previous results.

Further, we forecast that the degeneracy direction between \(\sigma_8\) and \(\Omega_\mathrm{m}\) given VSF data is roughly orthogonal to the posterior contours given weak lensing data, which is consistent with previous data analyses and forecasts \citep[e.g.,][]{Contarini2022EuclidCosmological,Contarini:2022mtu}. We do not explicitly show weak lensing posteriors, but rather show contours of constant \(S_8\) (the parameter combination most precisely constrained by weak lensing) to indicate the degeneracy direction. \citet{Preston:2025tyl} forecast that the combination of the Vera C. Rubin Observatory's Legacy Survey of Space and Time Year 1 weak lensing data (shear correlation function) and a prior on the strength of baryonic feedback, e.g., from measurements of the Sunyaev-Zeldovich effect in CMB experiments, can set a limit on the axion fraction that is comparable to the forecast \textit{Euclid} VSF limits. Such a future weak lensing dark matter model analysis requires an accurate model of the nonlinear matter power spectrum including feedback effects in non-cold dark matter models \citep{Vogt:2022bwy, Dome:2024hzq}. \citet{Schuster2024WhyCosmic} find that, unlike weak lensing, the void size function measured in upcoming surveys will be statistically insensitive to feedback. The linearity of voids, their robustness to astrophysical effects and orthogonal parameter degeneracies motivates future combined analyses of void statistics with galaxy clustering and weak lensing to break parameter degeneracies and set the strongest constraints on the nature of dark matter.

\subsection{Current approximations}
\label{sec:approx_discussion}

The forecast in \S~\ref{sec:voidforecast} makes the following approximations:
\begin{itemize}
    \item The mock void data come from a halo catalog at $z=0$, while a real survey spans a redshift range (Table \ref{tab:surveys-volume}).
    \begin{itemize}
        \item[-] The forecast thus neglects the redshift evolution of voids and information from the growth of structure over time. This approach means we are less sensitive to $\Omega_\mathrm{m}$ than previous \(\Lambda\)CDM forecasts \citep{Contarini2022EuclidCosmological}. For the axion forecast, we in any case use a CMB prior to constrain \(\Omega_\mathrm{m}\).
    \end{itemize}
    \item The minimum halo mass with which voids are defined is fixed to \(10^{13}\,M_\odot\) (\S~\ref{sec:HMF}).
    \begin{itemize}
        \item[-] This resolution limit means that we assume all the survey settings that we consider to have the same completeness down to \(10^{13}\,M_\odot\). This mass cutoff is a good approximation for the BOSS survey (especially in the CMASS redshift range $0.4 \lesssim z \lesssim 0.7$), but more recent surveys (e.g., DESI/\textit{Euclid}) may be sensitive to smaller halo masses. We anticipate that galaxy surveys sensitive to lower-mass halos will outperform our axion fraction forecast since these surveys will resolve smaller voids which are further suppressed in number density in the presence of axions (Figs.~\ref{fig:VSMF} and \ref{fig:vsf-model-ax-frac}).
    \end{itemize}
    \item The positions of tracers (halos) and voids are in real space, while an actual survey will measure the redshifts and angular positions of galaxies as tracers of voids.
    \begin{itemize}
        \item[-] This approximation is equivalent to the assumption that the void reconstruction procedure \citep[e.g.,][]{Nadathur:2018pjn,Nadathur2019BeyondBAO} exactly corrects for redshift-space distortions (RSD). Imperfections in the reconstruction will affect the results of the void-finding procedure (\S~\ref{sec:voidfinder}) and, in particular, the shapes of voids, although we do not use void ellipticity information in this work. Galaxy redshift-space distortions are a probe of ULAs \citep{Lague:2021frh,Rogers:2023ezo}. Neglecting galaxy RSD thus reduces the information with which we can constrain axions. Using galaxies as tracers also means that, in the model presented in \S~\ref{sec:VSF_model}, the halo bias is replaced by the galaxy bias. This change introduces a further sensitivity to how the halo-galaxy connection is modeled, e.g., by using a halo occupation distribution method \citep[e.g.,][]{Contarini2022EuclidCosmological}. We leave the study of the impact of galaxy bias and RSD on voids in the presence of axions (including void ellipticity) to future work.
    \end{itemize}
    \item We model ultra-light axions only through their effect on the linear matter power spectrum (\S~\ref{sec:VSF_model}) as traced by the void size function.
    \begin{itemize}
        \item[-] The forecast thus neglects the axion halo pressure effect \citep{Marsh:2013ywa,Winch:2024mrt}. Here, in addition to the suppression of matter fluctuations from which halos form, axion halos below a critical mass will not form at all, further suppressing the low-mass end of the halo mass function. (The halo mass function is not zero below this cutoff since halos will still form from the standard cold component of the dark matter, but their number density will be reduced.) This critical mass occurs when the average virial radius is less than the axion Jeans scale. \citet{Winch:2024mrt} demonstrate that this critical mass is less than the resolution limit in this work (\(10^{13} M_\odot\)) for all the axion particle masses we consider here. We anticipate that accounting for additional axion wave effects like these on smaller scales could increase sensitivity to axions and help to disentangle axion effects from other scale-dependent growth models (e.g., massive neutrinos). Nonetheless, for neutrinos, there is already a distinguishing effect on the background evolution \citep{Lesgourgues:2012uu}, which ULAs do not affect. It may be feasible to model axion wave effects in voids in a computationally-feasible way using the methods of \citet{Zimmermann:2024xvd,Zimmermann:2024vng}.
    \end{itemize}
\end{itemize}

\section{Conclusions} \label{sec:conclusion}

We investigate cosmic voids as a probe of the nature of dark matter for ongoing cosmological surveys, using ultra-light axions as a concrete model. Using mass-peak patch simulations with modified initial conditions (\S~\ref{sec:simulations}), we reproduce existing results on the effects of axions on the mass and spatial distribution of halos \citep{Lague:2021frh,Rogers:2023ezo}. First, as axion particle mass decreases, the Jeans matter suppression scale increases and so the low-mass end of the halo mass function is more suppressed (\S~\ref{sec:HMF}). Second, as axion particle mass decreases, low-mass halos become rarer and so the halo bias increases, boosting the amplitude of the halo-halo correlation function (\S~\ref{sec:axion_halo_bias}).

We use the \texttt{ZOBOV} method, as implemented in the \texttt{REVOLVER} code, to define voids as local underdensities of halos (\S~\ref{sec:voidfinder}). The primary effect of axions on voids is to remove low-mass halos, which leads to the merging of smaller (\(< 25\,\mathrm{Mpc}/h\)) voids into larger (\(> 25\,\mathrm{Mpc}/h\)) voids (Fig.~\ref{fig:Visualization}). This merging effect increases as particle mass decreases. The merging manifests in the void size function (number density as a function of radius) as a shift to larger radii (\S~\ref{subsec:voidsize}). We thus find that voids typically grow in size as particle mass decreases, but the voids get emptier of halos and fewer in number (Fig.~\ref{fig:single_void}). These effects suppress the overall amplitude of the void mass function (number density as a function of void mass as defined by the halos contained within; \S~\ref{subsec:voidsize}). The void-halo correlation function is a measure of the average number density of halos as a function of void radius (\S~\ref{subsec:voidcorr}). First, as axion mass decreases and the average void radius increases, the void-halo correlation shifts to larger separation. Second, as axion mass decreases and they get emptier of halos, the void-halo correlation gets more suppressed within the void, i.e., the halo number density profile of voids gets deeper. There is an equivalent effect (shifting and deepening) in the void-void correlation function, but the measurement is noisy owing to the smaller number of voids than halos in any given volume (\S~\ref{subsec:voidcorr}).

We forecast constraints on \(\Lambda\)CDM and axion parameters using the void size function as measured in three survey settings, which approximate the completed eBOSS survey \citep{Dawson2016eBOSS,Alam2017BOSS,Alam2021CompletedSDSS-IV} and the ongoing DESI \citep{Aghamousa2016DESI,AbdulKarim2025DESIDR2} and \textit{Euclid} \citep{euclid,Chudaykin2019MeasuringNeutrino,Contarini2022EuclidCosmological} experiments. We use our simulations as mock data with a Poisson likelihood to account for effective survey volume (\S~\ref{sec:VSF_likelihood}). We model the mock void size function data (\S~\ref{sec:VSF_model}) using an existing modification of the Sheth-Tormen formalism \citep{Sheth1999LargeScale,Sheth2001EllipsoidalCollapse,Sheth2004AHierarchy,Jennings2013TheAbundance,Contarini2022EuclidCosmological}. Axions enter the model through their small-scale suppression of the linear matter power spectrum. The linear halo bias is fit to the halo-halo correlation and the void punctual bias is fit to its maximum \(\Lambda\)CDM likelihood value to reproduce the planned survey analysis. We find that only with these linear bias parameters, we can fit well simulation results (Fig.~\ref{fig:vsf-with-data}), demonstrating how voids are much less sensitive to non-linearities than tracers of overdensities like galaxies. Nonetheless, we conclude that fitting the void punctual bias to \(\Lambda\)CDM simulations slightly biases inference on \(\Lambda\)CDM parameters in an axion model given the statistical precision of DESI and \textit{Euclid} data (Fig.~\ref{fig:validation-mcmc}). We therefore advocate that upcoming void size function analyses fully marginalize void bias parameters when inferring non-cold dark matter models. We find that the current approach otherwise slightly underestimates the strength of an axion signature.

A \textit{Euclid}-like survey, with a CMB \(\Lambda\)CDM prior from the Simons Observatory \citep{Ade2019SimonsObs,SimonsObservatory:2025wwn}, can improve the limit on the fraction of dark matter consisting of an axion with \(m_\mathrm{a} = 10^{-25}\,\mathrm{eV}\) by about a factor of two (\(< 4.6\%\) at 95\% credibility; \S\ref{sec:forecast_results}) over the current strongest bound from the Hubble galaxy UV luminosity function \citep[UVLF;][]{Winch:2024mrt}. A 10\% axion contribution can be recovered with a \(2\sigma\) preference with the same survey setting. A DESI-like survey achieves comparable results (Table \ref{tab:fax-bounds}). Further, in addition to being much less sensitive to non-linearities and baryonic feedback \citep{Schuster2024WhyCosmic} effects than other competitive probes, we demonstrate that axion effects in the void size function have a different degeneracy with \(\Lambda\)CDM parameters (Figs.~\ref{fig:combination-with-cmb} and \ref{fig:combination-with-WL}). We therefore advocate that future combined probes [e.g., with CMB lensing \citep{Hlozek2015ASearch,Dvorkin:2022bsc,Rogers:2023ezo}, galaxy weak lensing \citep{Dentler:2021zij,Preston:2025tyl}, Lyman-alpha forest \citep{Rogers:2023upm}, Webb galaxy UVLF \citep{Winch:2024mrt}] will be most powerful in disentangling the nature of dark matter. In future work, we will account for redshift-space distortions by modeling galaxies to define voids and extend this work to consider other deviations from the standard cold dark matter model (e.g., warm/interacting DM).

\begin{acknowledgments}
ASL was supported by an Aurora Borealis Fellowship from the Summer Undergraduate Research Program of the David A. Dunlap Department of Astronomy and Astrophysics at the University of Toronto. KKR is supported by an Ernest Rutherford Fellowship from the UKRI Science and Technology Facilities Council (grant no. ST/Z510191/1). The Dunlap Institute is funded through an endowment established by the David Dunlap family and the University of Toronto. RH acknowledges support from the Canadian Institute for Advanced Research, the Azrieli and the Alfred P. Sloan foundations and is supported by the Natural Sciences and Engineering Research Council of Canada Discovery Grant Program and the Connaught Fund. The authors at the University of Toronto acknowledge that the land on which the University of Toronto is built is the traditional territory of the Haudenosaunee and, most recently, the territory of the Mississaugas of the New Credit First Nation. They are grateful to have the opportunity to work in the community on this territory.
\end{acknowledgments}

%





\appendix
\label{appendix}
\section{Numerical convergence tests\label{app:simulation}}

\begin{figure}
    \centering
    \includegraphics[width=\linewidth]{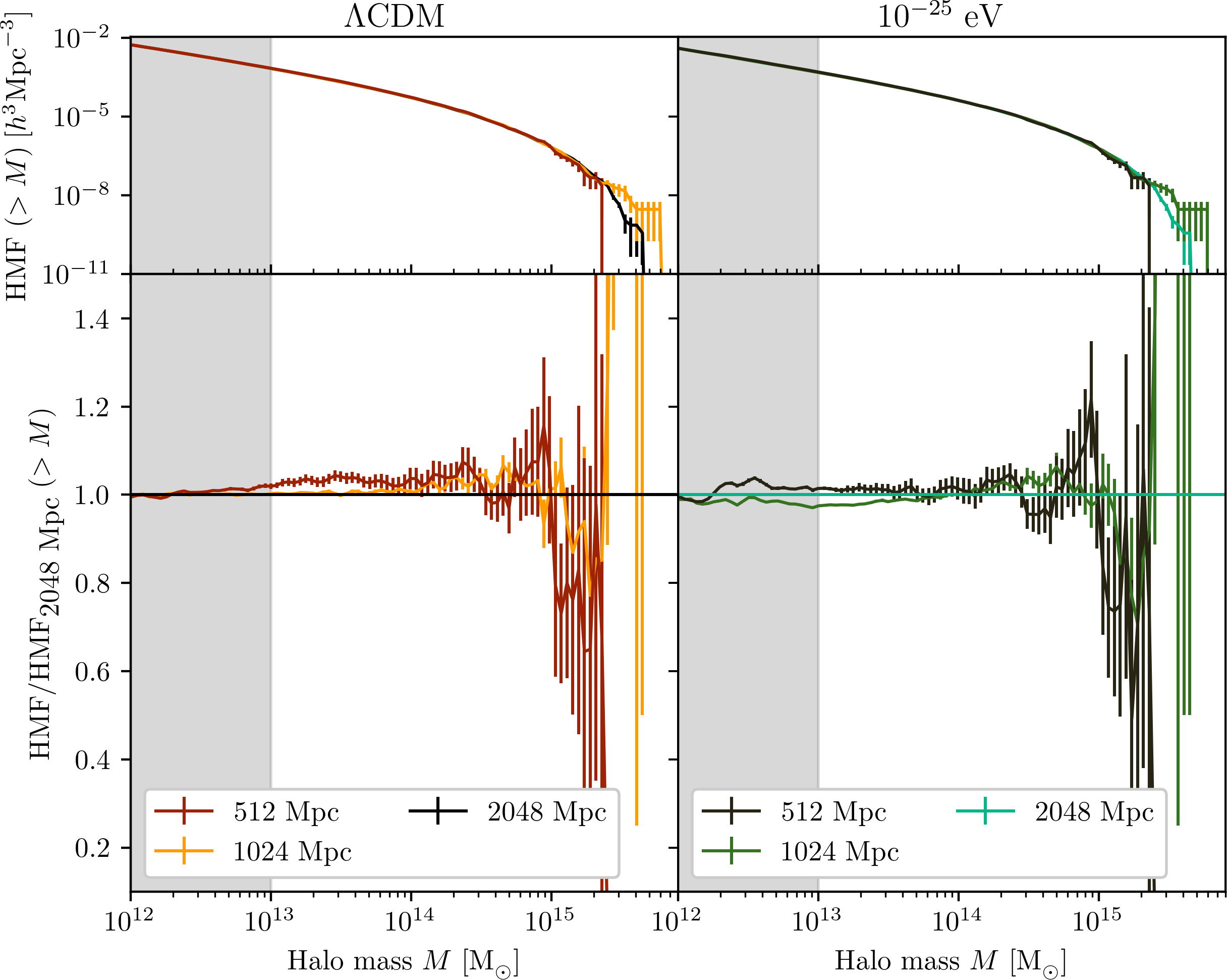}
    \caption{The halo mass function (HMF) of (\textit{left} panels) simulation \#1-3 and (\textit{right} panels) simulation \#4-6 (see Table \ref{parameters-table}), showing the effect of simulation volume while fixing the spatial resolution for both \(\Lambda\)CDM and axion (\(m_\mathrm{a} = 10^{-25}\,\mathrm{eV}\)) simulations. The \textit{upper} panels show the cumulative HMF (number density of halos \(>\) halo mass \(M\)). The \textit{bottom} panels show the ratio of the HMF to the case with the largest volume that we consider. We indicate the sample variance by the 68\% confidence error bars.}
    \label{fig:HMF_box_size}
\end{figure}

\begin{figure}
    \centering
    \includegraphics[width=0.5\linewidth]{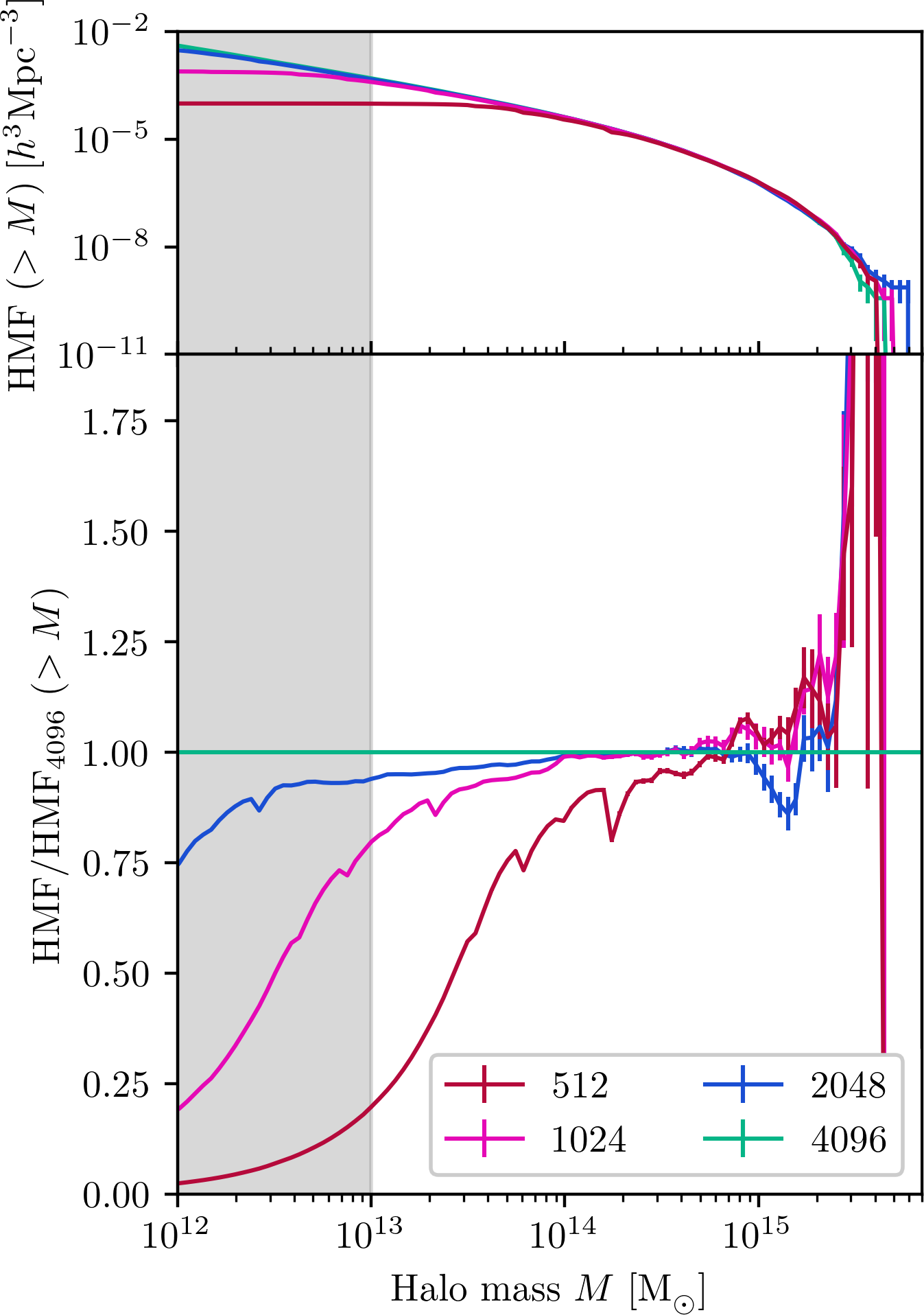}
    \caption{As Fig.~\ref{fig:HMF_box_size}, but showing the effect of spatial resolution while fixing the simulation volume for axion (\(m_\mathrm{a} = 10^{-25}\,\mathrm{eV}\)) simulations only (simulation \#6 and \#10-12; Table \ref{parameters-table}). We find that the simulations are more converged with respect to resolution if we remove all halos with \(M < 10^{13}\,\mathrm{M}_\odot\) (see also \S~\ref{sec:HMF}).}
    \label{fig:HMF_resolution}
\end{figure}

\begin{figure}
    \centering
    \includegraphics[width=\linewidth]{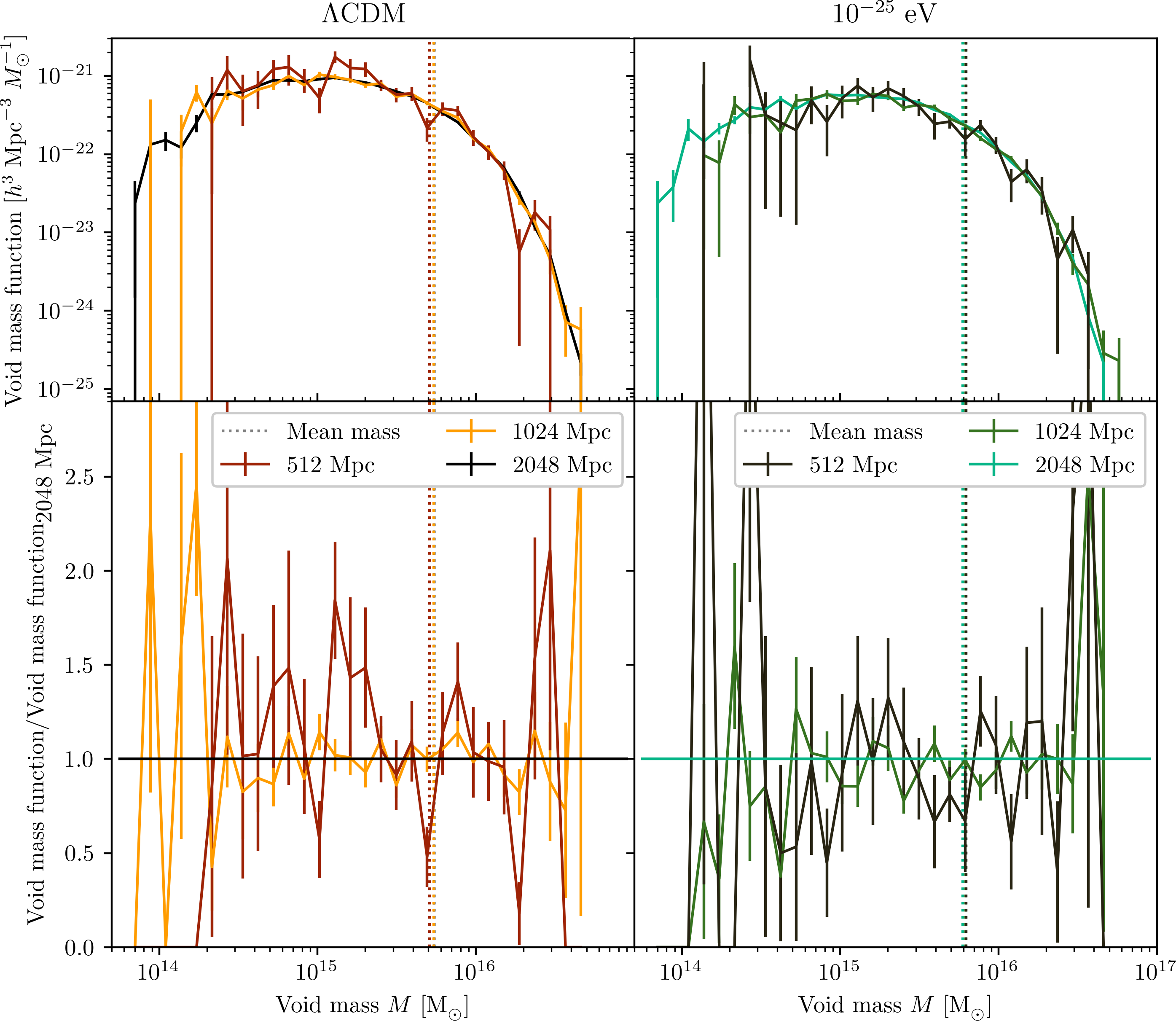}
    \caption{The void mass function (VMF) of (\textit{left} panels) simulation \#1-3 and (\textit{right} panels) simulation \#4-6 (see Table \ref{parameters-table}), showing the effect of simulation volume while fixing the spatial resolution for both \(\Lambda\)CDM and axion (\(m_\mathrm{a} = 10^{-25}\,\mathrm{eV}\)) simulations. The \textit{upper} panels show the VMF (number density of voids per unit void mass). The \textit{bottom} panels show the ratio of the VMF to the case with the largest volume that we consider. We indicate the sample variance by the 68\% confidence error bars and the average void mass by dotted lines.}
    \label{fig:VMF_box_size}
\end{figure}

\begin{figure}
    \centering \includegraphics[width=0.5\linewidth]{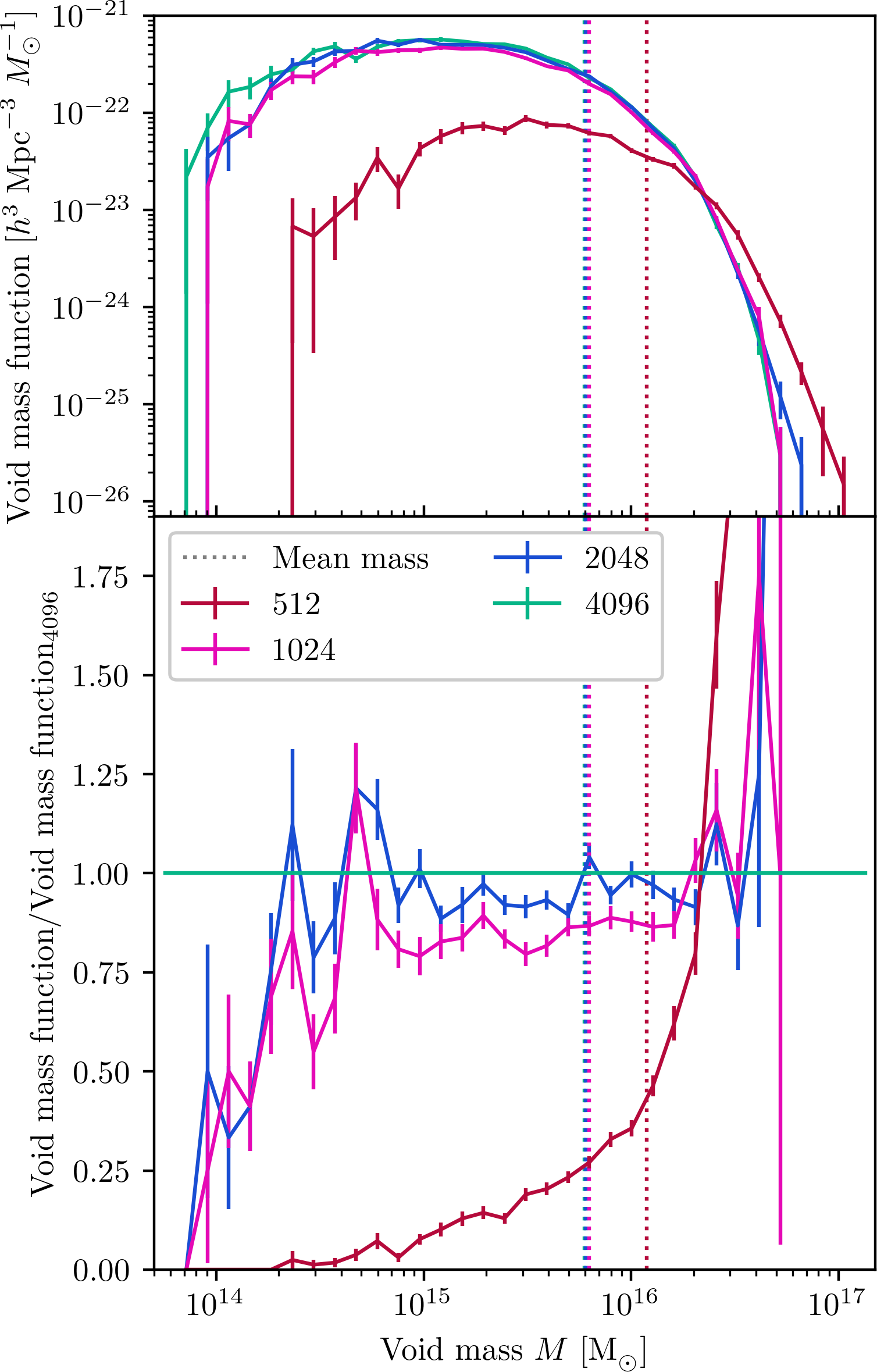}
    \caption{As Fig.~\ref{fig:VMF_box_size}, but showing the effect of spatial resolution while fixing the simulation volume for axion (\(m_\mathrm{a} = 10^{-25}\,\mathrm{eV}\)) simulations only (simulation \#6 and \#10-12; Table \ref{parameters-table}).}
    \label{fig:VMF_resolution}
\end{figure}

\begin{figure}
    \centering
    \includegraphics[width=\linewidth]{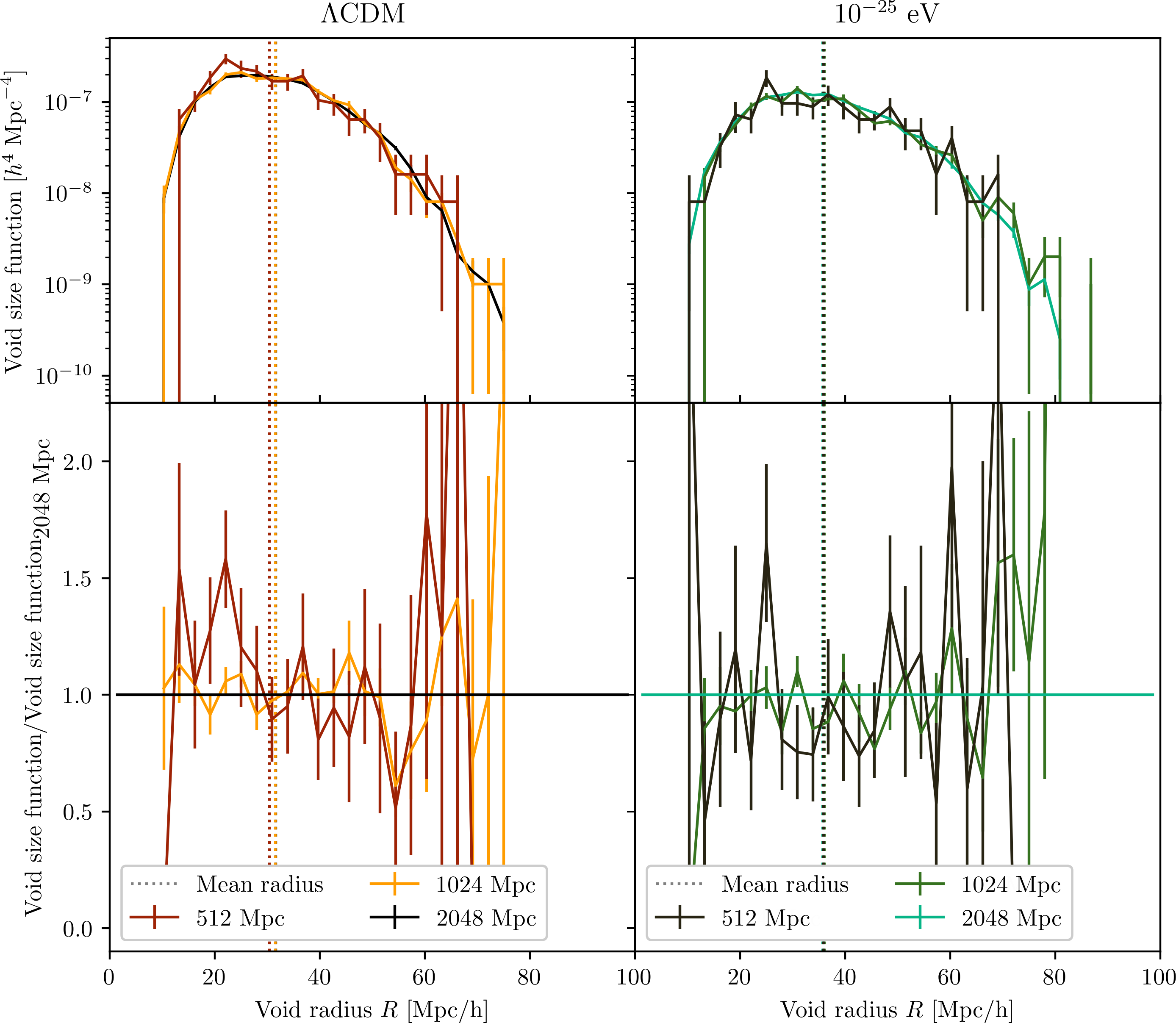}
    \caption{The void size function (VSF) of (\textit{left} panels) simulation \#1-3 and (\textit{right} panels) simulation \#4-6 (see Table \ref{parameters-table}), showing the effect of simulation volume while fixing the spatial resolution for both \(\Lambda\)CDM and axion (\(m_\mathrm{a} = 10^{-25}\,\mathrm{eV}\)) simulations. The \textit{upper} panels show the VSF (number density of voids per unit void radius). The \textit{bottom} panels show the ratio of the VSF to the case with the largest volume that we consider. We indicate the sample variance by the 68\% confidence error bars and the average void radius by dotted lines.}
    \label{fig:VSF_box_size}
\end{figure}

\begin{figure}
    \centering
    \includegraphics[width=0.5\linewidth]{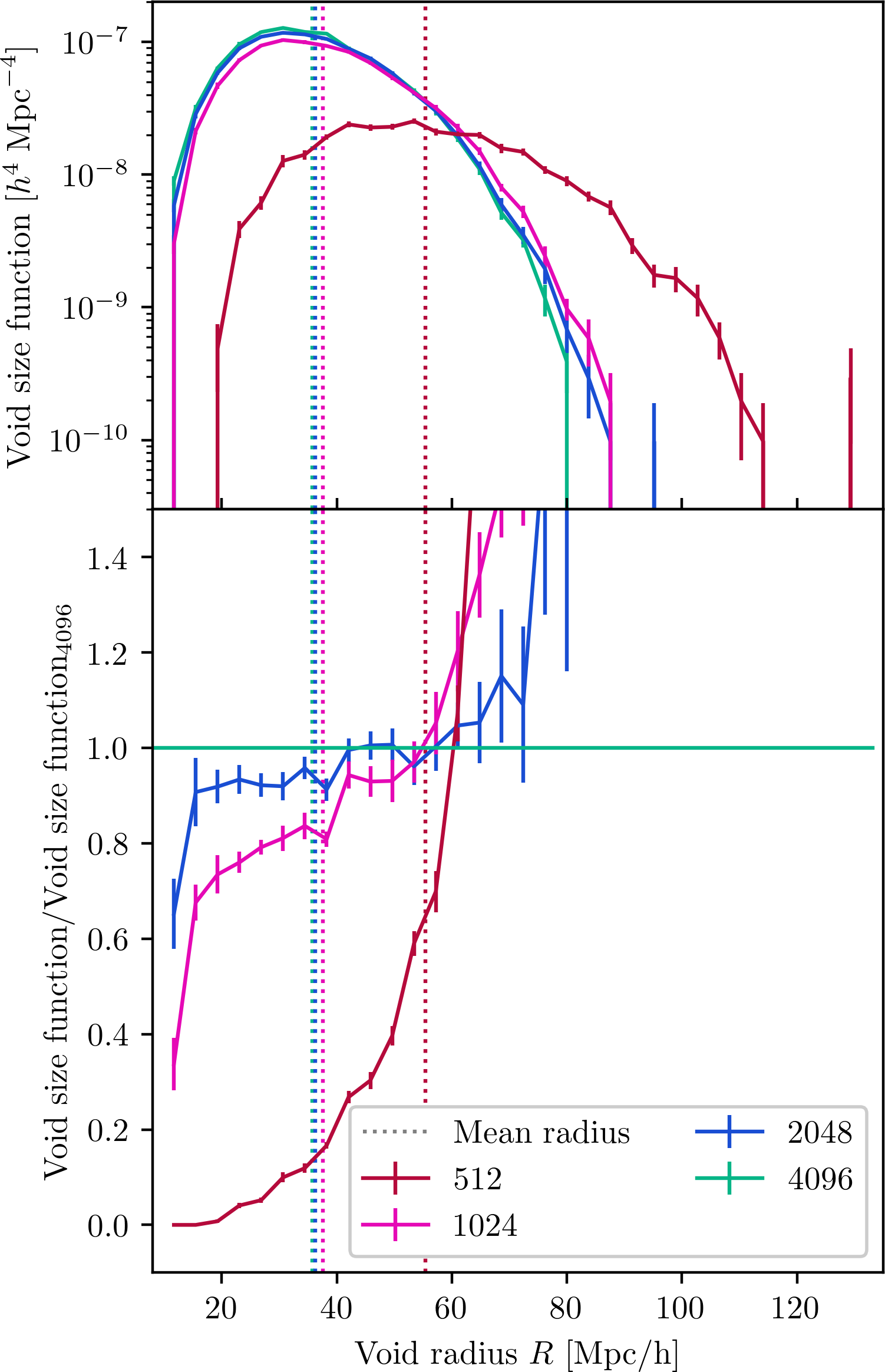}
    \caption{As Fig.~\ref{fig:VSF_box_size}, but showing the effect of spatial resolution while fixing the simulation volume for axion (\(m_\mathrm{a} = 10^{-25}\,\mathrm{eV}\)) simulations only (simulation \#6 and \#10-12; Table \ref{parameters-table}).}
    \label{fig:VSF_resolution}
\end{figure}

\begin{figure}
    \centering
    \includegraphics[width=\linewidth]{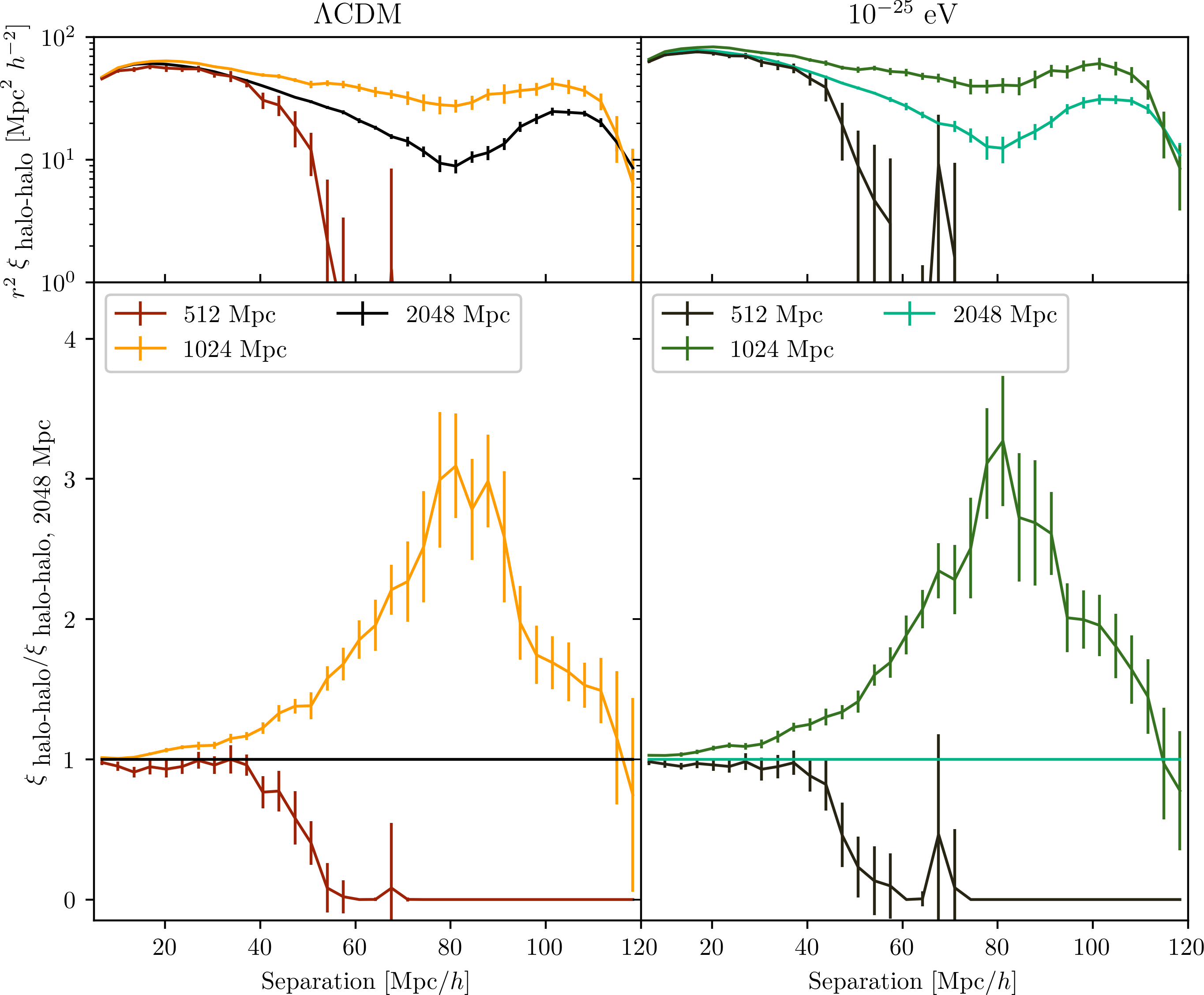}
    \caption{The halo-halo correlation function (HHCF) of (\textit{left} panels) simulation \#1-3 and (\textit{right} panels) simulation \#4-6 (see Table \ref{parameters-table}), showing the effect of simulation volume while fixing the spatial resolution for both \(\Lambda\)CDM and axion (\(m_\mathrm{a} = 10^{-25}\,\mathrm{eV}\)) simulations. The \textit{upper} panels show the HHCF; the \textit{bottom} panels show the ratio of the HHCF to the case with the largest volume that we consider. We indicate the sample variance by the 68\% confidence error bars. The HHCF is manifestly not converged, but the effect of simulation volume is the same for \(\Lambda\)CDM and axions, indicating that the relative effect of axions on the HHCF is converged with respect to simulation volume.}\label{fig:HHCF_box_size}
\end{figure}

\begin{figure}
    \centering
    \includegraphics[width=0.5\linewidth]{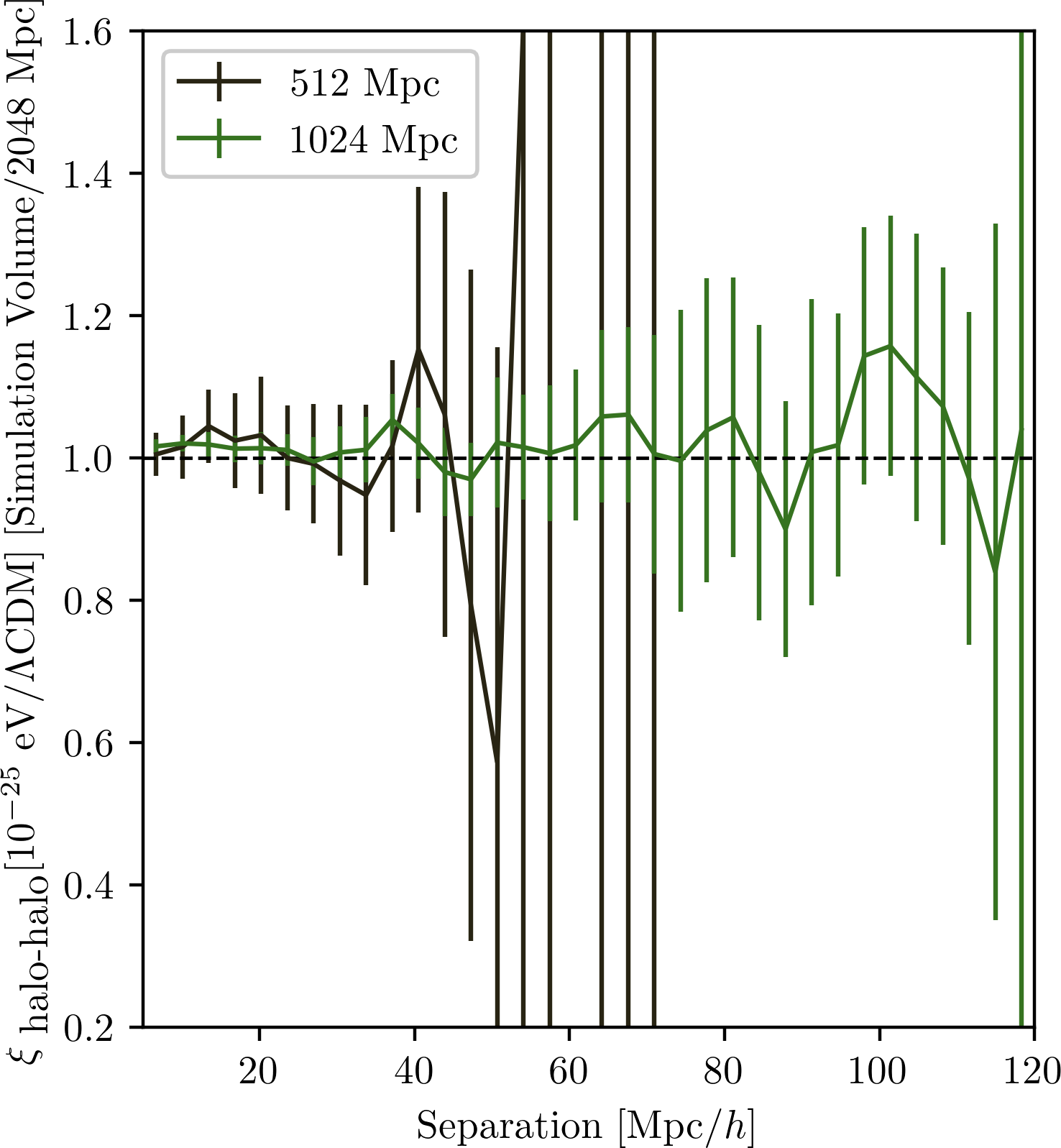}
    \caption{As Fig.~\ref{fig:HHCF_box_size}, but showing the ratio (\(10^{-25}\,\mathrm{eV} / \Lambda\mathrm{CDM}\)) of the ratios in the bottom panels (Simulation volume / 2048 Mpc). This quantity is consistent with unity indicating that the relative effect of axions on the HHCF is converged with respect to simulation volume.}\label{fig:HHCF_box_size_ratio}
\end{figure}

\begin{figure}
    \centering\includegraphics[width=0.5\linewidth]{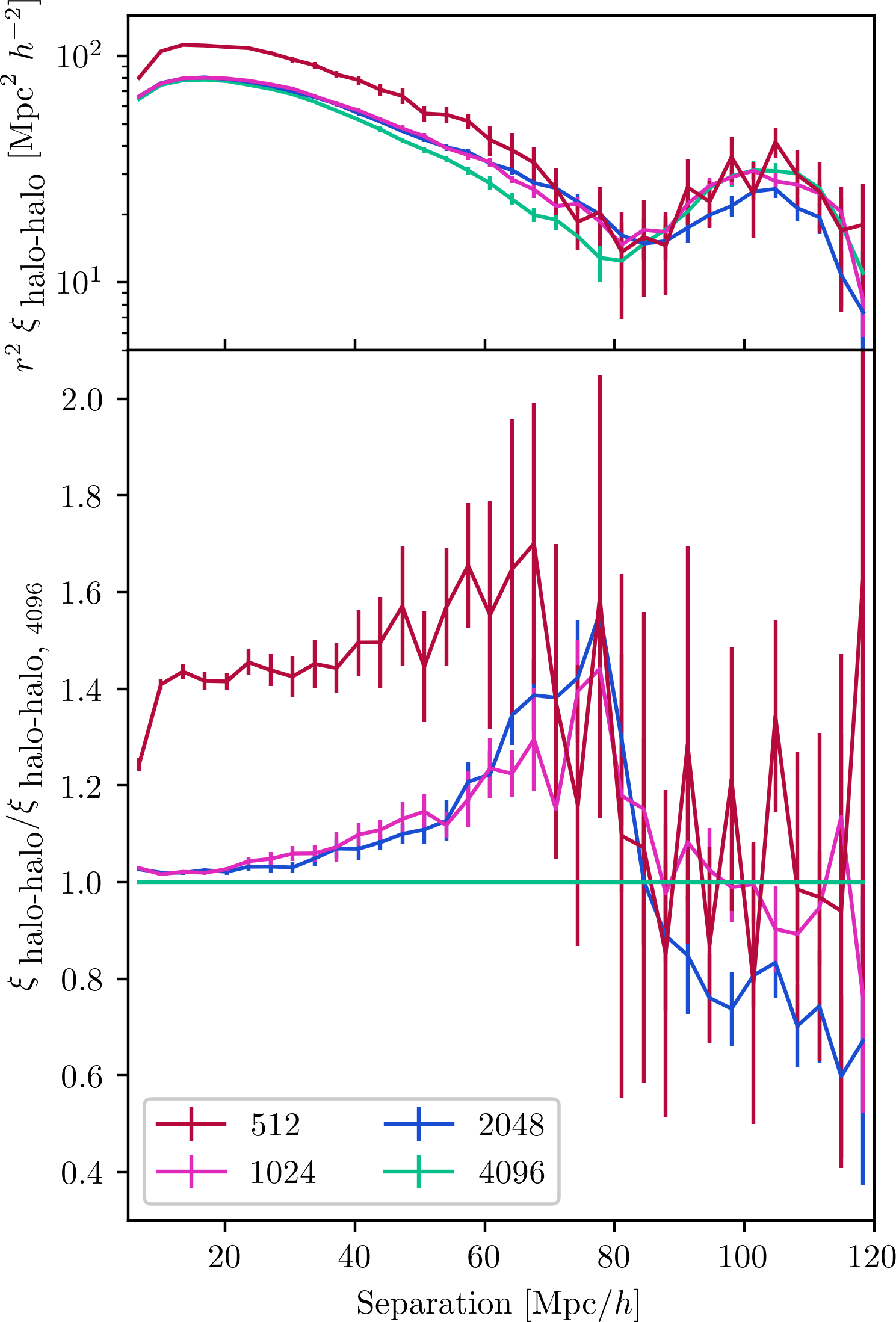}
    \caption{As Fig.~\ref{fig:HHCF_box_size}, but showing the effect of spatial resolution while fixing the simulation volume for axion (\(m_\mathrm{a} = 10^{-25}\,\mathrm{eV}\)) simulations only (simulation \#6 and \#10-12; Table \ref{parameters-table}).}\label{fig:HHCF_resolution}
\end{figure}

\begin{figure}
    \centering
    \includegraphics[width=\linewidth]{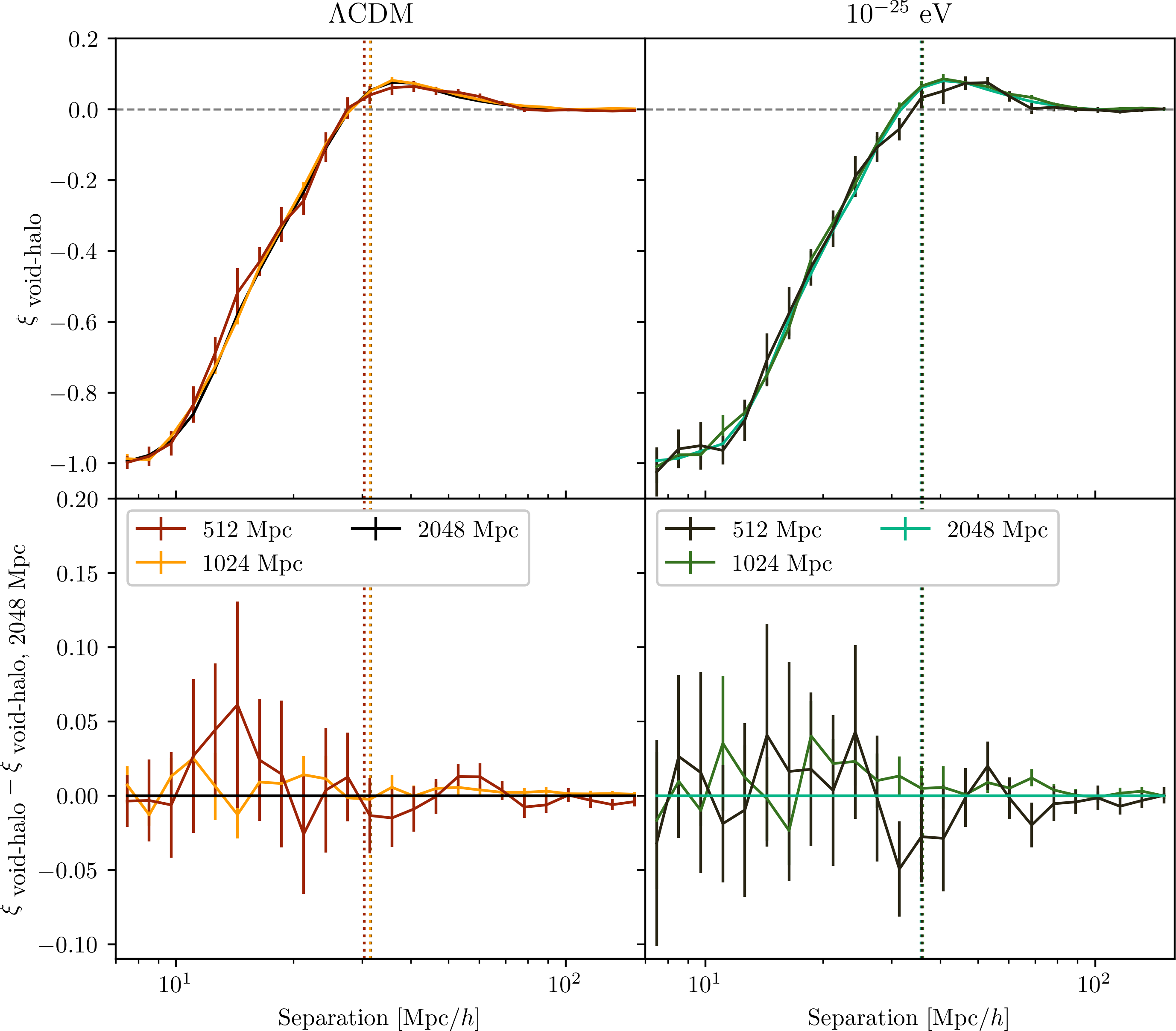}
    \caption{The void-halo correlation function (VHCF) of (\textit{left} panels) simulation \#1-3 and (\textit{right} panels) simulation \#4-6 (see Table \ref{parameters-table}), showing the effect of simulation volume while fixing the spatial resolution for both \(\Lambda\)CDM and axion (\(m_\mathrm{a} = 10^{-25}\,\mathrm{eV}\)) simulations. The \textit{upper} panels show the VHCF; the \textit{bottom} panels show the difference of the VHCF to the case with the largest volume that we consider. We indicate the sample variance by the 68\% confidence error bars and the average void radius by dotted lines.}\label{fig:VHCF_box_size}
\end{figure}

\begin{figure}
    \centering
    \includegraphics[width=0.5\linewidth]{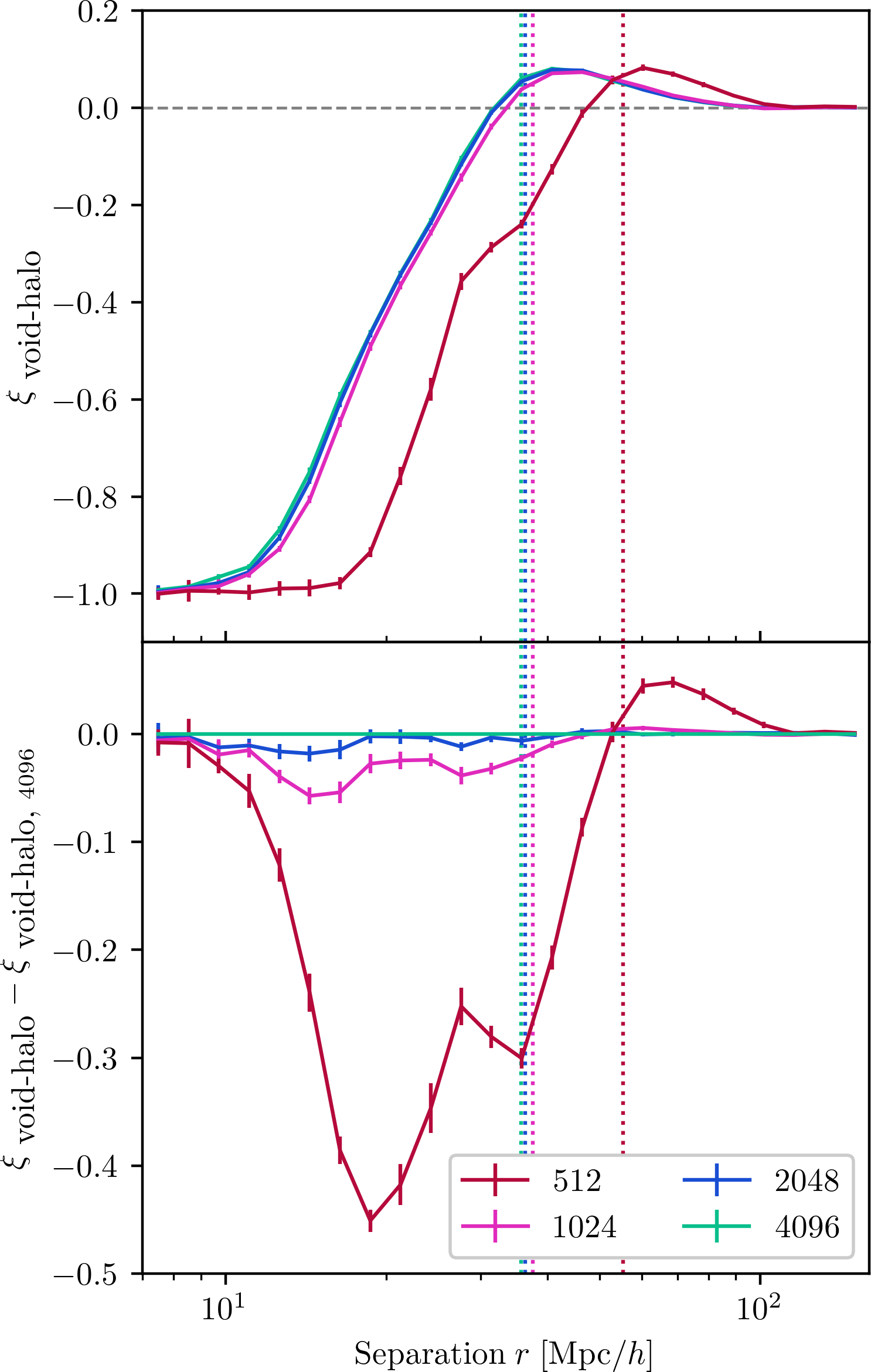}
    \caption{As Fig.~\ref{fig:VHCF_box_size}, but showing the effect of spatial resolution while fixing the simulation volume for axion (\(m_\mathrm{a} = 10^{-25}\,\mathrm{eV}\)) simulations only (simulation \#6 and \#10-12; Table \ref{parameters-table}).}\label{fig:VHCF_resolution}
\end{figure}

\begin{figure}
    \centering
    \includegraphics[width=\linewidth]{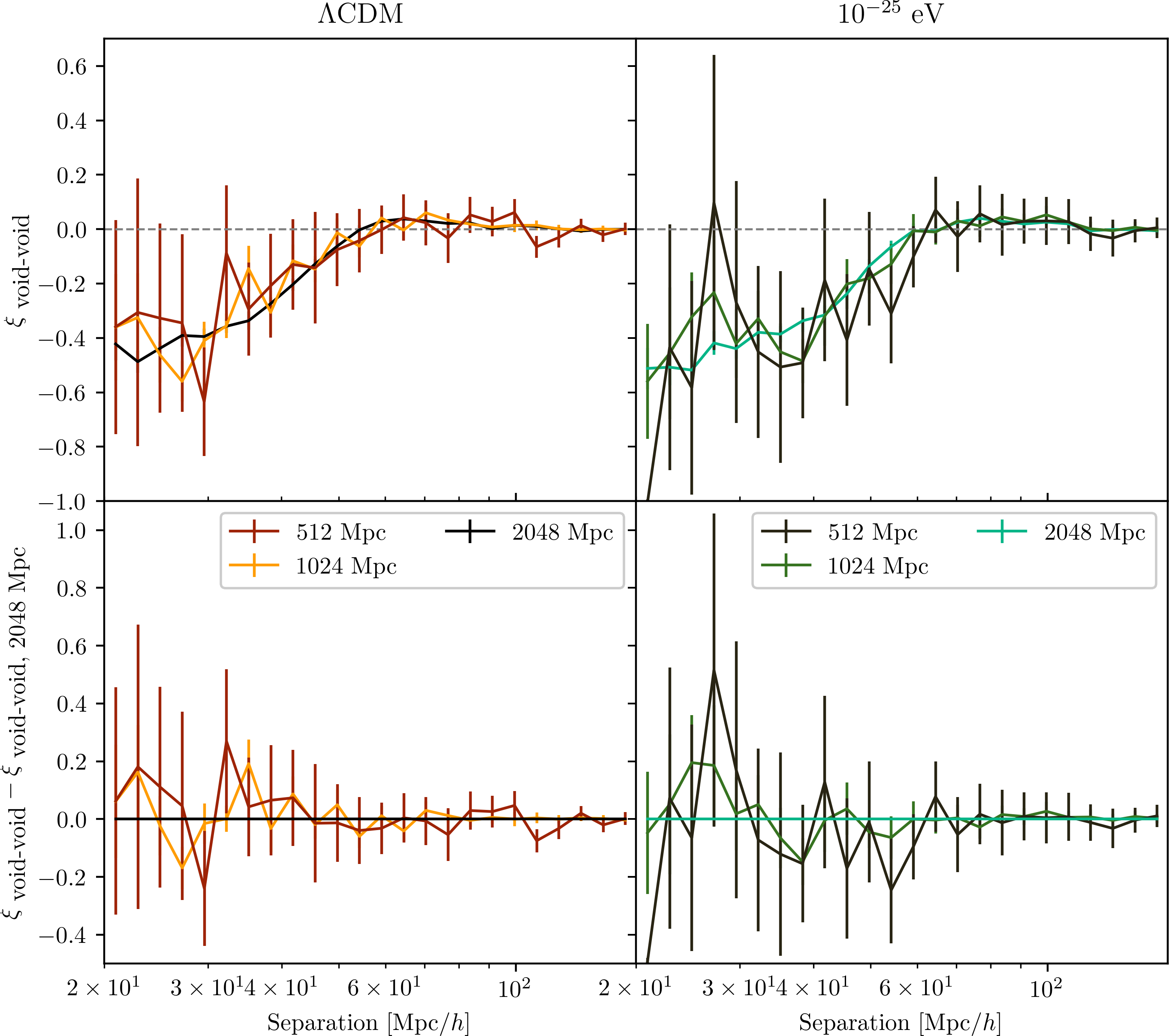}
    \caption{The void-void correlation function (VVCF) of (\textit{left} panels) simulation \#1-3 and (\textit{right} panels) simulation \#4-6 (see Table \ref{parameters-table}), showing the effect of simulation volume while fixing the spatial resolution for both \(\Lambda\)CDM and axion (\(m_\mathrm{a} = 10^{-25}\,\mathrm{eV}\)) simulations. The \textit{upper} panels show the VVCF; the \textit{bottom} panels show the difference of the VVCF to the case with the largest volume that we consider. We indicate the sample variance by the 68\% confidence error bars.}\label{fig:VVCF_box_size}
\end{figure}

\begin{figure}
    \centering
    \includegraphics[width=0.5\linewidth]{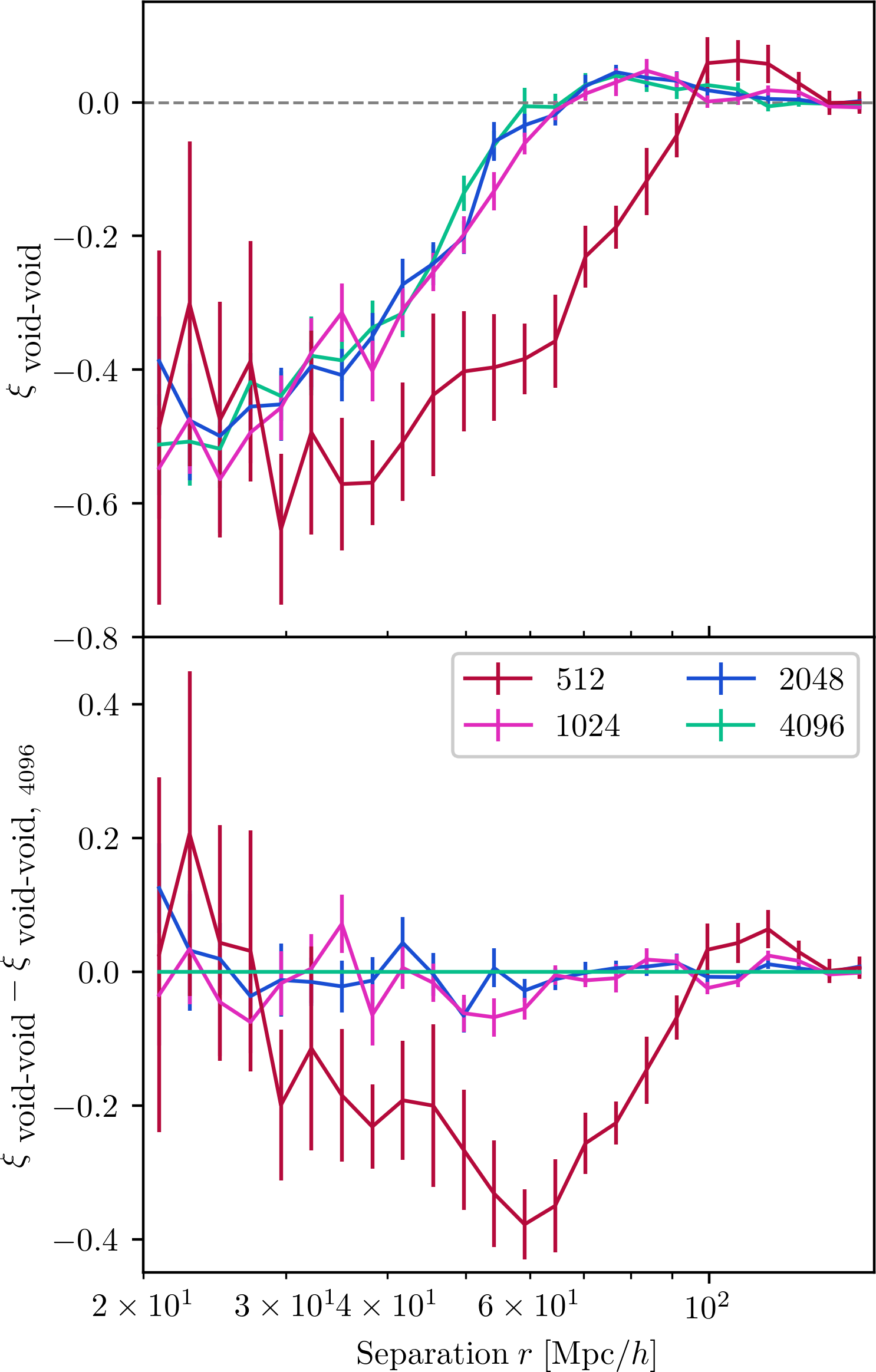}
    \caption{As Fig.~\ref{fig:VVCF_box_size}, but showing the effect of spatial resolution while fixing the simulation volume for axion (\(m_\mathrm{a} = 10^{-25}\,\mathrm{eV}\)) simulations only (simulation \#6 and \#10-12; Table \ref{parameters-table}).}\label{fig:VVCF_resolution}
\end{figure}

Figure \ref{fig:HMF_box_size} shows a numerical convergence test of the halo mass function as measured in \(\Lambda\)CDM and axion (\(m_\mathrm{a} = 10^{-25}\,\mathrm{eV}\)) simulations with respect to simulation volume while fixing spatial resolution. As volume increases, in particular, the most massive halos are better resolved since these derive from larger-scale density fluctuations. Fig.~\ref{fig:HMF_resolution} shows a numerical convergence test of the HMF with respect to spatial resolution while fixing simulation volume. As resolution increases, the least massive halos are better resolved since these derive from smaller-scale density fluctuations. Given this test, we remove all halos less massive than \(10^{13}\,\mathrm{M}_\odot\), which are not well resolved even in our baseline volume [\((2048\,\mathrm{Mpc})^3\)] and resolution (\(4096^3\) cells).

Figure \ref{fig:VMF_box_size} shows a numerical convergence test of the void mass function with respect to simulation volume while fixing spatial resolution. As volume increases, both the \(\Lambda\)CDM and axion simulation voids are well converged with the noisiness of the measurement decreasing as more voids in a larger volume are resolved. Fig.~\ref{fig:VMF_resolution} shows a numerical convergence test of the VMF with respect to spatial resolution while fixing simulation volume. As resolution increases, the least massive voids are better resolved since these are defined by the structure formed from lower-mass halos, which are themselves better resolved. Resolving lower-mass voids means that spuriously-massive voids are split up into more lower-mass voids. Figs.~\ref{fig:VSF_box_size} and \ref{fig:VSF_resolution} show the equivalent convergence tests but for the void size function. The trends are equivalent with more massive voids equating to larger voids and less massive voids equating to smaller voids.

Figure \ref{fig:HHCF_box_size} shows a numerical convergence test of the halo-halo correlation function with respect to simulation volume while fixing spatial resolution. The HHCF is manifestly not converged on the largest scales with respect to simulation volume. We checked that this remains true as we vary the minimum and maximum halo masses in the catalog. We hypothesize that this lack of convergence arises from a masking effect. Increasing simulation volume resolves larger, more massive halos (Fig.~\ref{fig:HMF_box_size}). Because the mass peak-patch simulations do not permit overlapping halos (\S~\ref{sec:simulations}), resolving larger halos effectively masks the larger number of smaller halos that otherwise would manifest. Although larger halos are exponentially fewer, their masking effect changes the correlation structure of the more numerous smaller halos, hence the large effect on the HHCF. We leave detailed studies of this effect to future work. However, we note that the effect of simulation volume is the same for the \(\Lambda\)CDM and axion catalogs. We explicitly illustrate this fact in Fig.~\ref{fig:HHCF_box_size_ratio}, where we show that the ratio (\(10^{-25}\,\mathrm{eV} / \Lambda\mathrm{CDM}\)) of the ratios in the bottom panels of Fig.~\ref{fig:HHCF_box_size} (Simulation volume / 2048 Mpc) are consistent with unity. This means that the \textit{relative} effect of axions on the HHCF, i.e., the boost in halo bias, is consistently captured as we vary simulation volume. Fig.~\ref{fig:HHCF_resolution} shows a numerical convergence test of the HHCF with respect to spatial resolution while fixing simulation volume. As resolution increases, the smaller-scale correlation structure of halos is better resolved.

Figure \ref{fig:VHCF_box_size} shows a numerical convergence test of the void-halo correlation function with respect to simulation volume while fixing spatial resolution. As volume increases, both the \(\Lambda\)CDM and axion simulations are well converged with the noisiness of the measurement decreasing as more objects (halos and voids) in a larger volume are resolved. Fig.~\ref{fig:VHCF_resolution} shows a numerical convergence test of the VHCF with respect to spatial resolution while fixing simulation volume. As resolution increases, smaller voids are better resolved, shifting the VHCF to smaller separations. Figs.~\ref{fig:VVCF_box_size} and \ref{fig:VVCF_resolution} show the equivalent convergence tests but for the void-void correlation function. The trends are equivalent.


\bibliography{sample631}{}
\bibliographystyle{aasjournal}



\end{document}